\documentclass[10pt,twocolumn]{iopart}

\usepackage{citesort}
\usepackage{iopams}
\usepackage{graphicx}
\usepackage{setstack}

\begin{document}

\title[]{Influence of edge magnetization and electric fields on zigzag silicene, germanene and stanene nanoribbons}
\author{Ayami Hattori$^1$, Keiji Yada$^1$, Masaaki Araidai$^{2,3}$, Masatoshi Sato$^{4}$, Kenji Shiraishi$^2$, Yukio Tanaka$^1$}
\address{$^1$ Department of Applied Physics, Nagoya University, Nagoya 464-8603, Japan}
\address{$^2$ Institute of Materials and Systems for Sustainability, 
Nagoya University, Nagoya 464-8603, Japan}
\address{$^3$ Institute for Advanced Research, Nagoya University, Nagoya 464-8601, Japan}
\address{$^4$ Yukawa Institute for Theoretical Physics, Kyoto University, Kyoto 606-8502, Japan}
\ead{hattori@rover.nuap.nagoya-u.ac.jp} 

\begin{abstract}
Using a multi-orbital tight-binding model, we have studied the edge states of zigzag silicene, germanene, and stanene nanoribbons (ZSiNRs, ZGeNRs and ZSnNRs, respectively) in the presence of the Coulomb interaction and a vertical electric field. The resulting edge states have non-linear energy dispersions due to multi-orbital effects, and the nanoribbons show induced magnetization at the edges. 
Owing to this non-linear dispersion, ZSiNRs, ZGeNRs and ZSnNRs may not provide superior performance in field effect transistors, as has been proposed from single-orbital tight-binding model calculations. 
We propose an effective low-energy model that describes the edge states of ZSiNRs, ZGeNRs, and ZSnNRs. 
We demonstrate that the edge states of ZGeNR and ZSnNR show anti-crossing of bands with opposite spins, even if only out-of-plane edge magnetization is present. The ability to tune the spin polarizations of the edge states by applying an electric field points to future opportunities to fabricate silicene, germanene and stanene nanoribbons as spintronics devices.
\end{abstract}

\vspace{2pc}
\noindent{\it Keywords\/}: edge states, multi-orbital model, edge magnetization, electric field effect, non-linear dispersion, mixed spin states \par
\submitto{\JPCM} 
\noindent (Some figures may appear in colour only in the online journal)

\maketitle
\ioptwocol

\section{Introduction} 
Silicene, germanene and stanene are novel two-dimensional honeycomb allotropes containing silicon, germanium and tin atoms, respectively. They are elemental analogues of graphene, which show rich and exotic electronic properties. 
Epitaxial growth of silicene and germanene on metallic substrates and of stanene on a Bi$_2$Se$_3$ substrate has been successfully demonstrates  \cite{Ag3,Ag2,Zr,Au,Pt,Bi2Te3}, and planar stanene has been grown on a Ag substrate \cite{Yuhara}. 
Recently, several groups have tried to fabricate silicene, germanene, and stanene on insulating substrates \cite{gerMoS2,Araidai}. Silicene and germanene have also attracted much attention from the context of the chemical modification \cite{germanane}. 
From their low dimensionality, these materials are expected to have applications in nanoelectronic devices with atomic layer thicknesses \cite{siligerreview,structure_sil,sildevice,structure_ger,ZrB22016,multilayersil}. 

Silicene, germanene and stanene nanoribbons with zigzag edges, which we call zigzag silicene nanoribbons (ZSiNRs), zigzag germanene nanoribbons (ZGeNRs), zigzag stanene nanoribbons (ZSnNRs), respectively, support helical edge states in the bulk energy gaps of their nonmagnetic states due to their large spin-orbit coupling. The helical edge states comprise two counter-propagating edge modes that carry only a spin current along an edge of the sample \cite{QSHjpsj}. In contrast, in the presence of Coulomb interactions, the single-orbital Kane-Mele-Hubbard model has been used to demonstrate that the helical edge states show ferromagnetism and that their properties change drastically, depending on the ferromagnetic configuration \cite{kanemelehubbard}. In particular, zigzag nanoribbons with large spin-orbit coupling exhibit large magnetic anisotropy, which can be used to produce devices that control spin channels. 

In contrast, these three materials have low-buckled structures and they host $sp^3$-like hybridized orbitals rather than $sp^2$ orbitals \cite{takedashiraishi}. The low-buckled structure also makes it possible to control the energy gap by applying an electric field, which can be used to design a field-effect transistor composed of silicene, germanene  or stanene \cite{silgerQSH,electric,Ezawa_externalfield}.  Preliminary experiments aimed at producing such a field-effect transistor using silicene have already begun \cite{sildevice}. 
The approximation of the single-orbital tight-binding model also show that vertical electric fields can easily control the helical edge states of ZSiNRs, ZGeNRs and ZSnNRs, based on which a topological field-effect transistor has been proposed \cite{Ezawa_FET}.

Our previous study, however, has shown that the actual energy dispersion of the edge states of ZSiNR, ZGeNR, and ZSnNR is highly non-linear due to the low-buckled geometry \cite{edge_states}. This important feature cannot be captured by the single-orbital tight-binding model because it has accidental chiral symmetry \cite{Ezawa_externalfield}. To evaluate these nanoribbons for real applications, we therefore need to consider multi-orbital effects, instead of using the simplest single-orbital model. 

In this study, we use a multi-orbital tight-binding model to study edge states in the presence of the Coulomb potential and a vertical electric field. The multi-orbital model can consider the low-buckled geometry. We consider both out-of-plane antiparallel edge magnetization (OP-AFM) and out-of-plane parallel edge magnetization (OP-FM) \cite{spinfilter,fujitager,systematic_stanene}. We calculate the energy spectra of the edge states for nonmagnetic states, OP-AFM and OP-FM with and without vertical electric fields. To understand the complicated behaviors of the edge states, we also derive an effective low-energy model that describes the non-linear dispersion of the edge states of ZSiNR, ZGeNR and ZSnNR, and we demonstrate that the edge states of ZGeNR and ZSnNR show anti-crossing of bands with opposite spins due to edge magnetization and spin-orbit coupling. We also show that it will be difficult to create a field-effect transistor using ZSiNR, ZGeNR or ZSnNR because of the complicated non-linear dispersion of the edge states. However, our results also imply that these nanoribbons can be used to create spintronics devices. 

The organization of this paper is as follows. In \sref{sec2}, we first explain the structures of nanoribbons with two types of spin configurations and formulations. We introduce a multi-orbital Hubbard model to calculate the induced magnetization at the edge. In \sref{sec3}, we show the results of our numerical calculations of the multi-orbital tight-binding model for the nonmagnetic, OP-AFM and OP-FM states without and with vertical electric fields. We find an effective Hamiltonian for the edge states that explains the non-linear dispersion and spin polarization. In \sref{sec4}, we summarize our results.

\section{Models and formulations}\label{sec2}
\subsection{Atomic structures of nanoribbons and edge magnetization configuration}
\begin{figure}[htbp] 
\includegraphics[width=8cm]{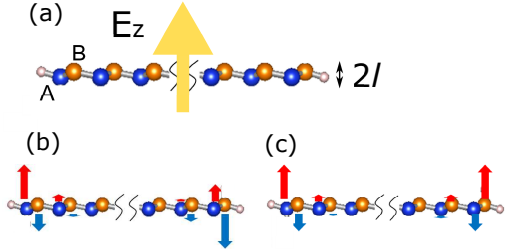}
\caption{\label{fig:3states} (color online) (a) Side view of a 1H/1H ZNR. $E_z$ is a perpendicular electric field, and 2$l$ is the height between the A (blue sphere)  and B (orange sphere) sublattice. (b) and (c) Illustration of, respectively, out-of-plane and antiparallel edge magnetization (OP-AFM) and out-of-plane and parallel edge magnetization (OP-FM) in a 1H/1H ZNR. The red (blue) arrow denotes up (down) spin. }
\end{figure}

In our previous work \cite{edge_states}, we have assumed that zigzag nanoribbons with low-buckled geometries (ZNRs) are terminated by hydrogen. In the present paper, we calculate the following two cases: mono-hydrogen termination at both edge sites (1H/1H) or di-hydrogen termination at both edge sites (2H/2H). In our model, unit cell contains two atoms, and we denote the width of a nanoribbon by $w$; we choose $w=100$ ($\sim$0.1$\mu$m) in the following. We calculate the energy spectra of edge states with and without applying an electric field perpendicular to the nanoribbon. We consider nonmagnetic states, with either out-of-plane and antiparallel edge magnetization (OP-AFM) or out-of-plane and parallel edge magnetization (OP-FM) (\Fref{fig:3states}) by determining the edge magnetization self-consistently. 

\subsection{Multi-orbital Hubbard model}
The multi-orbital Hubbard model is defined by the following Hamiltonian, 
\begin{eqnarray}
\mathcal{H}_{Hub} = \mathcal{H}_0+\mathcal{H}_{\mathrm{so}}+\mathcal{H}_{\mathrm{H}} + \mathcal{H}_{E_z} + \mathcal{H}_U
\end{eqnarray}
The first term, $\mathcal{H}_0$, expresses the on-site energy of the silicon, germanium or tin atoms (tetragens) and includes nearest neighbor hopping between them: 
\begin{eqnarray}
\mathcal{H}_0=&\sum_{\langle i,j \rangle} \sum_{\alpha,\beta} \sum_{\tau} (t^{\alpha\beta}_{i,j} c^\dagger_{i\alpha \tau}c_{j\beta \tau}+\mathrm{h.c.}) \nonumber \\
&+\sum_{i} \sum_{\alpha} \sum_{\tau} \epsilon_\alpha c^\dagger_{i\alpha\tau}c_{i\alpha\tau},\label{eq7}
\end{eqnarray}
where $c^\dagger_{i\alpha \tau}$ and $c_{i\alpha \tau}$ are the creation and annihilation operators for an electron in atomic orbital $\alpha$ and with a spin $\tau$ at site $i$. The quantity $\epsilon_{\alpha}$ denotes the site energy for orbital $\alpha$. The first term in equation \eref{eq7} corresponds to  hybridization between the tetragens, and the second term represents the on-site energy at the tetragen sites. The indices $\langle i,j \rangle$ run over all the nearest neighbor hopping sites. The hopping integral $t^{\alpha\beta}_{i,j}$ is determined by the Slater--Koster parameters as shown in \cite{slater}. 
The second term $\mathcal{H}_{\mathrm{so}}$ expresses the spin--orbit interaction 
\begin{equation}
\mathcal{H}_{\mathrm{so}}=\frac{\xi_0}{2}\sum_{i} \sum_{\bar{\alpha}\bar{\beta}\bar{\gamma}} \sum_{\tau,\tau^{\prime}} \epsilon_{\bar{\alpha}\bar{\beta}\bar{\gamma}}c^{\dagger}_{i\bar{\alpha} \tau}(-i\hat{\sigma}_{\bar{\gamma}}) c_{i\bar{\beta} \tau^{\prime}}+\mathrm{h.c.}, 
\end{equation} 
where $\xi_0$ is the strength of the spin--orbit coupling, $\bar{\alpha}=x,y,z $, $\bar{\beta}=x,y,z$ and $\bar{\gamma}=x,y,z$ are indices of the $p_{\bar{\alpha}}$, $p_{\bar{\beta}}$ and $p_{\bar{\gamma}}$ orbitals. The quantity $\epsilon_{\bar{\alpha}\bar{\beta}\bar{\gamma}}$ is an antisymmetric tensor and 
$\hat{\sigma}_{\bar{\gamma}}$ is the Pauli matrix acting on the spin space.
The third term, $\mathcal{H}_\mathrm{H}$, describes the hydrogen termination: 
\begin{eqnarray}
\mathcal{H}_\mathrm{H}=&\sum_{\langle i,j \rangle} \sum_\alpha \sum_{\tau} (t^{s\alpha}_{ij} d^{\dagger}_{is\tau} c_{j\alpha\tau}+\mathrm{h.c.}) \nonumber \\
&+\sum_i \sum_{\tau} \epsilon_H d^{\dagger}_{is\tau} d_{is\tau},\label{eq9}
\end{eqnarray}
where $d^\dagger_{is\tau}$ and $d_{is\tau}$ are the creation and annihilation operators for an electron at hydrogen site $i$. The first term in equation \eref{eq9} corresponds to hybridization of the hydrogen and the tetragen, and the second term is the on-site energy at a hydrogen site, where $\epsilon_{H}$ denotes the site energy for an electron at a hydrogen atom. The parameters in $\mathcal{H}_0$, $\mathcal{H}_\mathrm{so}$ and $\mathcal{H}_\mathrm{H}$ are adopted from our previous work \cite{edge_states}. 
The fourth term, $\mathcal{H}_{E_z}$, describes electrostatic potential. 
\begin{eqnarray}
\mathcal{H}_{E_z} = -l E_z \sum_{i,\mu} \mu_i c^\dagger_{i\alpha\tau} c_{i\alpha\tau},
\end{eqnarray}
where $l$ (\AA) represents the buckling height and $2l$ is height between the A and B sublattices, as shown in \fref{fig:3states}. We assume that the values of $l$ for silicene, germanene and stanene are $0.23$ \AA, $0.33$ \AA, and $0.4$ \AA \mbox{ }\cite{Ezawa_review} and --that $\mu_i=+1(-1)$ for an A (B) sublattice site.-- The fifth term, $\mathcal{H}_{U}$, expresses the Coulomb interaction. 
\begin{eqnarray}
\mathcal{H}_{U} = \sum_{i,\alpha} U_{\alpha} 
n_{i,\alpha,\uparrow} n_{i,\alpha,\downarrow}, 
\end{eqnarray}
where $\alpha$ and $U_\alpha$ are the orbital index and the magnitude of the on-site Coulomb potential of orbital $\alpha$ and where $n_{i,\alpha,\tau}=c^\dagger_{i,\alpha,\tau}c_{i,\alpha,\tau}$. We implement the mean-field approximation in our multi-orbital tight-binding model and write $\mathcal{H}_U$ in the form  
\begin{eqnarray}
\mathcal{H}_U = 
&\sum_{i} U_{p_z} \left( \langle n_{i,p_z,\downarrow} \rangle n_{i,p_z,\uparrow}
+\langle n_{i,p_z,\uparrow} \rangle n_{i,p_z,\downarrow}
\right) \\
&+\sum_{i} U_{s} \left( \langle n_{i,s,\downarrow} \rangle n_{i,s,\uparrow}
+\langle n_{i,s,\uparrow} \rangle n_{i,s,\downarrow}
\right). 
\end{eqnarray}

We compare the energy dispersions of the edge states of the ZNRs with and without spin-orbit coupling using first-principles calculations with our multi-orbital model \cite{hydrosil_nm,functionalized_germanene,systematic_stanene}. The magnetization configurations are shown in figures \ref{fig:3states}(b) and (c). We consider two types of configurations: (1) out-of-plane and antiparallel edge magnetization (OP-AFM) and  (2) out-of-plane and parallel edge magnetization (OP-FM). The magnitude of the magnetization becomes a maximum at the edge site, and it decays with oscillations from the edge sites into the bulk. 
We only take into account $U_{p_z}$ in ZSiNR and ZSnNR, since the effects of the edge sites are mainly contributed from the $p_z$ orbital. For ZGeNR, we consider both $U_{p_z}$ and $U_s$, since the contribution of the s-orbital is also prominent. We use $U_{p_z} = 2.4 $ eV for ZSiNR, $U_s = 2.5 $ eV and $U_{p_z} = 1.8 $ eV for ZGeNR, and $U_{p_z} = 1.5$ eV for ZSnNR. The energy dispersions based on these parameter choices are consistent with first-principles calculations \cite{hydrosil_nm,functionalized_germanene,systematic_stanene}.

\section{Results}\label{sec3}
In this section, we denote the energy dispersions of edge states with up spins at a right edge (red line), those with down spins at a right edge (yellow line), those with up spins at a left edge (blue line) and those with down spins at a left edge (green line) by $E_{R,\uparrow,\mathrm{state}}$, $E_{R,\downarrow,\mathrm{state}}$, $E_{L,\uparrow,\mathrm{state}}$, and $E_{L,\downarrow,\mathrm{state}}$, respectively. 
Here the `state' in the subscript denotes either a nonmagnetic case (nm), an OP-AFM case (af) or an OP-FM case (fm). The direction of the magnetization for OP-AFM is opposite at the left and right edges, while that for OP-FM is the same. (see Appendix figures \ref{sil_mag}-\ref{sta_mag_k})

\subsection{Edge magnetization} 
\begin{figure*}[htbp]
\includegraphics[width=18cm]{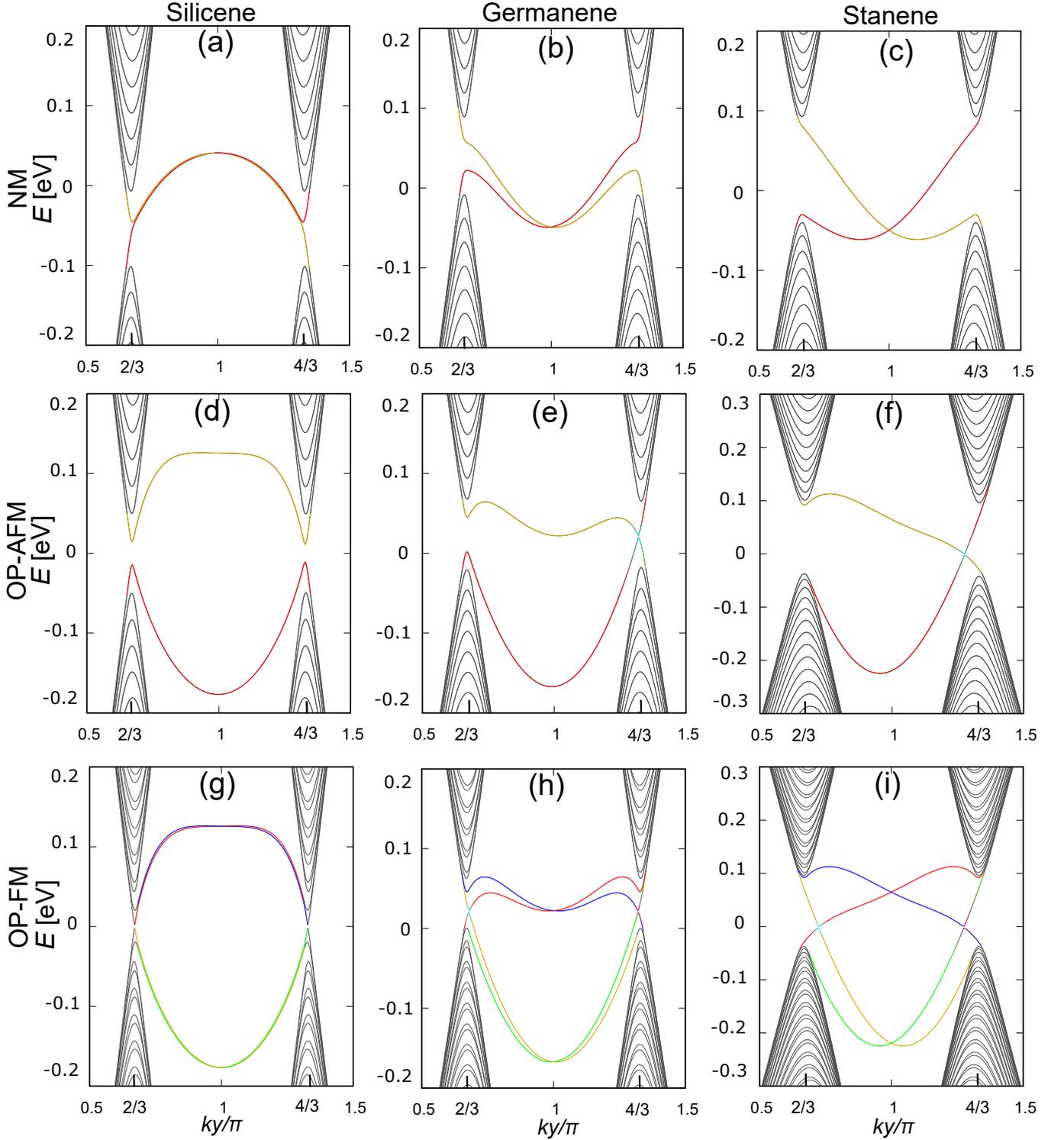}
\caption{\label{nofield} (color online) Energy spectra of 1H/1H nonmagnetic, OP-AFM and OP-FM ZSiNRs [(a), (d) and (g), respectively], ZGeNRs [(b), (e) and (h)] and ZSnNRs [(c), (f) and (i)] for w=100. The value $E=0$ represents Fermi energy. Red, yellow, blue and green denote the $E_{R,\uparrow,\mathrm{state}}$, $E_{R,\downarrow,\mathrm{state}}$, $E_{L,\uparrow,\mathrm{state}}$, and $E_{L,\downarrow,\mathrm{state}}$ of the ZNRs, respectively. For OP-AFM ZGeNR (e) and ZSnNR (f), the edge states show anti-crossing of bands with opposite spins and the color of $E_{R,\uparrow,\mathrm{af}}$ gradually changes from red to yellow around the energy gap near $k_y=4\pi/3$ while that of $E_{R,\downarrow,\mathrm{af}}$ gradually changes from yellow to red. Also, for OP-FM ZGeNR (h) and ZSnNR (i), $E_{R,\uparrow,\mathrm{fm}}$ and $E_{R,\downarrow,\mathrm{fm}}$ [$E_{L,\uparrow,\mathrm{fm}}$ and $E_{L,\downarrow,\mathrm{fm}}$] show anti-crossing of bands with opposite spins near $k_y=2\pi/3$ ($4\pi/3$). }
\end{figure*}

We show the energy dispersions of the edge states of nonmagnetic, OP-AFM and OP-FM 1H/1H ZSiNRs in figures \ref{nofield}(a), (d), and (g); 1H/1H ZGeNRs in figures \ref{nofield}(b), (e) and (h) and 1H/1H ZSnNRs in figures \ref{nofield}(c), (f) and (i). 

The nonmagnetic ZNRs in figures \ref{nofield}(a)-(c) exhibit helical edge states, which appear in the bulk energy gap. The edge states connect the valence and conduction bands: the helical edge states $E_{R,\uparrow,\mathrm{nm}}$ (red) connect the valence band at $k_y=2\pi/3$ and the conduction band at $k_y=4\pi/3$, while $E_{R,\downarrow,\mathrm{nm}}$ (yellow) connects the conduction band at $k_y=2\pi/3$ and the valence band at $k_y=4\pi/3$. The edge states $E_{R,\uparrow,\mathrm{nm}}$ and $E_{L,\downarrow,\mathrm{nm}}$ (green) [$E_{R,\downarrow,\mathrm{nm}}$ and $E_{L,\uparrow,\mathrm{nm}}$ (blue)] are degenerate. In \fref{nofield}(a), except for $k_y=\pi$, ZSiNR exhibits a small spin splitting, due to the weak spin-orbit coupling of a silicon atom and it shows upward-convex dispersion. In contrast, ZGeNR [\fref{nofield}(b)] and ZSnNR [\fref{nofield}(c)] display larger spin splitting than ZSiNR and show downward-convex dispersion. Their dispersions are non-linear because of the low-buckled geometry; note that these dispersions are different from those calculated using the single-orbital tight-binding model \cite{Ezawa_externalfield}. 

Next, we consider the energy spectra of the OP-AFM 1H/1H ZNRs shown in figures \ref{nofield}(d)-(f). 
Although the magnetization breaks both time-reversal and inversion symmetry, the combined symmetry that includes both time-reversal and inversion operation remains. 
In this case, $E_{R,\uparrow,\mathrm{af}}$ (red) and $E_{L,\downarrow,\mathrm{af}}$ (green) [$E_{R,\downarrow,\mathrm{af}}$ (yellow) and $E_{L,\uparrow,\mathrm{af}}$ (blue)] are doubly degenerate. 
\Fref{nofield}(d) shows that the helical edge states of ZSiNR disappear and that the energy gap opens due to the edge magnetization in comparison with the nonmagnetic ZSiNR shown in \fref{nofield}(a). 
The energy gap at $k_y=2\pi/3$ is similar in size to that at $k_y=4\pi/3$.  
In contrast, the energy dispersion of OP-AFM 1H/1H ZGeNR has an energy gap at $k_y=2\pi/3$ that is similar to ZSiNR, and it almost crosses at $k_y=4\pi/3$, as shown in \fref{nofield}(e). According to Kane-Mele-Hubbard model, the energy dispersion of an OP-AFM ZNR crosses near the $K$ point \cite{kanemelehubbard}. However, in the OP-AFM ZGeNR calculations based on the multi-orbital model, the edge magnetization and the on-site spin-orbit coupling of the germanium atoms mix bands with opposite spins, and an energy gap opens slightly near $k_y=4\pi/3$. Here, the edge states $E_{R,\uparrow,\mathrm{af}}$ and $E_{R,\downarrow,\mathrm{af}}$ show anti-crossing of bands with opposite spins. 
\Fref{nofield}(e) shows that the color of $E_{R,\uparrow,\mathrm{af}}$ changes gradually from red to yellow near $k_y=4\pi/3$, as it goes from left to right along $k_y$, while that of $E_{R,\downarrow,\mathrm{af}}$ also changes gradually from yellow to red. 
This shows that both up and down spins flip around this point. The edge states around the energy gap are called `mixed spin states (channels)' \cite{graMoS2hetero}. 
The energy spectrum of the OP-AFM 1H/1H ZSnNR shown in \fref{nofield}(f) has a similar behavior to that of ZGeNR. 

\begin{figure}[htbp]
\includegraphics[width=8cm]{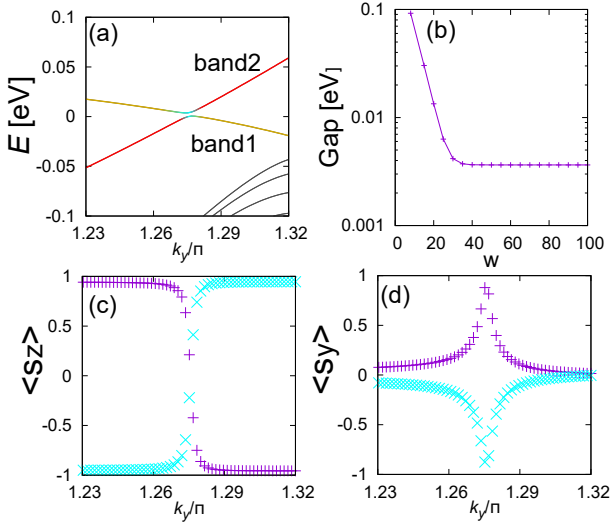}
\caption{\label{invert} (color online) (a) Energy spectra of OP-AFM ZSnNR [enlarged from \fref{nofield}(f)] (b) Energy gap as a function of ribbon width $w$. (c) and (d): Expectation values of $\langle s_z\rangle$ and $\langle s_y\rangle$, the $z$ and $y$ components, respectively, of the momentum-decomposed spin for band1 (purple) and band2 (sky blue).}
\end{figure}

Let us consider in detail the mixed spin states of 1H/1H OP-AFM ZSnNR. \Fref{invert}(a) shows the energy spectra of the edge states of the OP-AFM ZSnNR near $k_y=4\pi/3$ according to our multi-orbital tight-binding model [enlarged from \fref{nofield}(f))]. \Fref{invert}(b) shows the energy gap between band1 and band2 of \fref{invert}(a) as a function of the ribbon width $w$. 
The energy gap becomes constant ($E_\mathrm{gap} \sim 0.0036$ eV) when $w$ is larger than 30, which means that the energy gap is not due to the finite-size effect. 
Figures \ref{invert}(c) and (d) show the expectation values $\langle s_z\rangle$ and $\langle s_y\rangle$ of the $z$ and $y$ components, respectively, of the momentum-decomposed spin of band1 and band2. 
When $\langle s_z\rangle$ inverts near the energy gap, the magnitude of $\langle s_y\rangle$ becomes a maximum and the edge states have spin components of in-plane direction (the $y$-direction). This verifies that the mixed-spin states cause the energy gap to open and the spins around the energy gap to flip. 

Third, we show the energy spectra of the OP-FM 1H/1H ZNRs. 
Here, time-reversal symmetry is broken by the magnetization, but the inversion symmetry remains.  
As a result, the OP-FM ZNRs exhibit dispersion symmetric with respect to $k_y=\pi$. In contrast to the OP-AFM ZNRs, only the direction of magnetization at the right-side edge is reversed (see figures \ref{sil_mag}, \ref{ger_mag}, \ref{sta_mag}). As a result, $E_{L,\uparrow,\mathrm{fm}}(k_y)$ (blue) and $E_{L,\downarrow,\mathrm{fm}}(k_y)$ (green) in figures \ref{nofield}(g)-(i) coincide with $E_{L,\uparrow,\mathrm{af}}(k_y)$ (blue) and $E_{L,\downarrow,\mathrm{af}}(k_y)$ (green) in figures \ref{nofield}(d)-(f).  
On the other hand, $E_{R,\uparrow,\mathrm{fm}}(k_y)$ (red) and $E_{R,\downarrow,\mathrm{fm}}(k_y)$ (yellow) in figures \ref{nofield}(g)-(i) coincide with $E_{R,\downarrow,\mathrm{af}}(2\pi-k_y)$  (yellow) and $E_{R,\uparrow,\mathrm{af}}(2\pi-k_y)$ (red). 
The edge states of ZSiNR have small energy gaps near $k_y=2\pi/3$ and $4\pi/3$, as shown in \fref{nofield}(g). 
On the other hand, for ZGeNR [\fref{nofield}(h)] and ZSnNR [\fref{nofield}(i)], $E_{R,\uparrow,\mathrm{fm}}$ and $E_{R,\downarrow,\mathrm{fm}}$ show anti-crossing of bands with opposite spins near $k_y=2\pi/3$; note that $E_{L,\uparrow,\mathrm{fm}}$ and $E_{L,\downarrow,\mathrm{fm}}$ also show anti-crossing of bands with opposite spins near $k_y=4\pi/3$. Then, as shown in figures \ref{nofield}(h) and (i), the color of $E_{R,\uparrow,\mathrm{fm}}$ gradually changes from red to yellow near $k_y=2\pi/3$ and that of $E_{R,\downarrow,\mathrm{fm}}$ also gradually changes from yellow to red. Also, the spins of $E_{L,\uparrow,\mathrm{fm}}$  and $E_{L,\downarrow,\mathrm{fm}}$ flip near $k_y=4\pi/3$.

The energy spectra of the nonmagnetic, OP-AFM and OP-FM 2H/2H ZNRs are shown in \fref{nofield_k} (Appendix) and are similar to those of the 1H/1H ZNRs. 


Finally, in order to understand in more detail the energy dispersion of the edge states of 1H/1H ZNRs with edge magnetization, we introduce the following effective low-energy model :  
\begin{eqnarray}
\mathcal{H}_\mathrm{eff}=\mathcal{H}_\mathrm{nm}
+\mathcal{H}_\mathrm{m}, \label{eff}
\end{eqnarray}
with
\begin{eqnarray}
\mathcal{H}_\mathrm{nm} &= (a_0+a_1 k^2)\sigma_0\tau_0 + vk\sigma_z\tau_z \\ 
\mathcal{H}_\mathrm{m} &= (m_{0} + m_{1}k^2) \sigma_z\tau_i,
\end{eqnarray}
,where $a_0$, $a_1$, $v$, $m_{0}$ and $m_{1}$ are constants determined by fitting the energy dispersion of our multi-orbital tight-binding model near $k_y=\pi$. $\sigma_i$ is a Pauli matrix in spin space and $\tau_0$ and $\tau_z$ are Pauli matrices that distinguish the left and right edge sites of the ZNRs. Here, $i=z$ (OP-AFM) or $i=0$ (OP-FM). $\mathcal{H}_\mathrm{nm}$ describes the energy dispersion of the nonmagnetic states of the ZNRs, and the solution is $\epsilon_\mathrm{nm}(k) = a_0+a_1 k^2\pm vk$. In the single-orbital model, $a_1$ is zero, since the energy dispersion near $k_y=\pi$ is always linear. However, in practice, the second-order term in $k$ (with coefficient $a_1$) must be included to describe the non-linear dispersion of the ZNRs due to chiral symmetry breaking, as shown in \tref{fitpara}. The quantity $v$ is the strength of the spin-orbit coupling. The magnitude of $v$ is large for ZSnNR, while that for ZSiNR is small. $\mathcal{H}_\mathrm{m}$ describes the edge magnetization, which breaks time-reversal symmetry and cause an energy gap  to open near $k_y=\pi$ The solutions are $\epsilon_\mathrm{afm}(k) = a_0+a_1k^2\pm (vk+m_0+m_1k^2)$ for OP-AFM and $\epsilon_\mathrm{fm}(k) = a_0+a_1k^2\pm (vk+m_0+m_1k^2)$ and $a_0+a_1k^2\pm (vk-m_0-m_1k^2)$ for OP-FM. The crossing point, which is located at $k_y=\pi$ in the nonmagnetic ZNRs, moves away from $k_y=\pi$ in the OP-AFM ZNRs, as shown in figures \ref{nofield}(e) and (f). As shown in \tref{fitpara}, $\mathcal{H}_\mathrm{m}$ also requires the second-order term in $k$ (with coefficient $m_1$), since the edge states do not exhibit symmetric dispersion with respect to $E=0$, which is different from the single-orbital tight-binding model \cite{kanemelehubbard}. 

However, the mixed spin states of the edge states cannot be explained by $\mathcal{H}_\mathrm{m}$. This term causes the energy gap of the edge states to open at the crossing point in OP-AFM ZGeNR and ZSnNR and in OP-FM ZGeNR and ZSnNR, as shown in figures \ref{nofield}(e), (f), (h) and (i).  
On the basis of $\langle s_y \rangle $ shown in \fref{invert}(d), we introduce the additional Hamiltonian $\mathcal{H}_\Delta = m_y \sigma_y\tau_0$ to the effective model; it can qualitatively explain the mixed spin states. 
In particular, $m_y$ is significant for materials with a large magnitude of spin-orbit coupling.

\begin{table*}
\caption{\label{fitpara} Parameters of the effective Hamiltonian $\mathcal{H}_\mathrm{eff}$ obtained by fitting to the multi-orbital tight-binding model for ZNRs near $k_y=\pi$.} 

\begin{indented}
\lineup
\item[]\begin{tabular}{@{}*6{c}} 
\br 
System      &  $a_0 $ [eV]  &  $a_1$ [eV/\AA$^2$]  &  $v$ [eV/\AA] & $m_0$ [eV]  &  $m_1$ [eV/\AA$^2$]  \cr
\mr
Silicene (nm)     & $0.0420$ & -0.751&  $0.0109$ &  &    \cr 
\mr
Silicene (afm/fm)    & $-0.0256$      & 1.11  & $0.00931$& $-0.152$         &  $0.257$   \cr
\mr
Germanene (nm) & $-0.0497$         & 1.76   &$0.0620$&           &    \cr
\mr
Germanene (afm/fm) & $-0.0722$   &  0.956 &$0.0321$& $-0.0954$   & $-0.0290$   \cr
\mr
Stanene (nm)    & $-0.0506$        & 0.139      &$0.229$&           &    \cr
\mr
Stanene (afm/fm)    & $-0.0758$  & 2.87       &$0.236$& $-0.146$  & $3.75$    \cr
\br
\end{tabular}
\end{indented}
\end{table*}

\subsection{Vertical electric field effect}
In this section, we calculate the energy spectra of nonmagnetic, OP-AFM and OP-FM 1H/1H and 2H/2H ZNRs in the presence of a uniform electric field perpendicular to the plane of the ZNRs. If we apply such an electric field to nonmagnetic bulk silicene, germanene or stanene, the magnitude of the energy gap at the $K$ point decreases, and the gap closes completely at a critical electric field, $E_\mathrm{cr}$. If we apply an electric field larger than $E_\mathrm{cr}$, the energy gap starts to open again. The critical electric field $E_\mathrm{cr}$ differs depending on the material. We find $E^{(\mathrm{Si})}_{\mathrm{cr}}= 8.87$ [meV/\AA], $E^{(\mathrm{Ge})}_{\mathrm{cr}}= 85.2$ [meV/\AA] and  $E^{(\mathrm{Sn})}_{\mathrm{cr}}= 161$ [meV/\AA], and we have calculated the energy spectra for 1H/1H ZSiNR, ZGeNR and ZSnNR ($w=100$) at $2E_\mathrm{cr}$.  

\begin{figure*}[htbp]
\includegraphics[width=18cm]{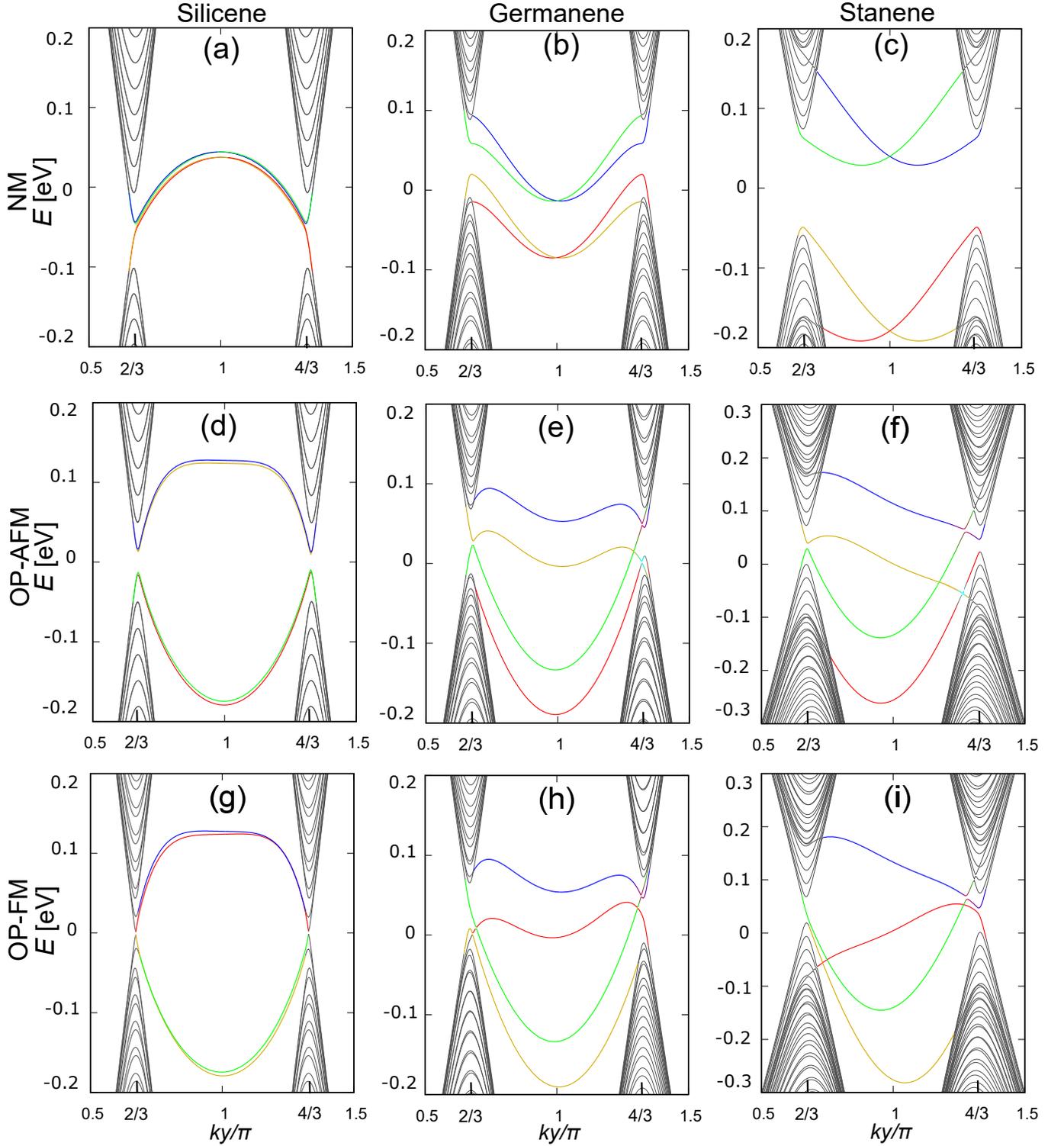}
\caption{\label{field} (color online) Energy spectra of 1H/1H nonmagnetic, OP-AFM and OP-FM ZSiNRs [(a), (d) and (g), respectively]; ZGeNRs [(b), (e) and (h)] and ZSnNRs [(c), (f) and (i)] in the presence of a vertical electric field $E_z=2E_\mathrm{cr}$ for $w=100$. The value $E=0$ represents the Fermi energy. Red, yellow, blue and green denote the $E_{R,\uparrow,\mathrm{state}}$, $E_{R,\downarrow,\mathrm{state}}$, $E_{L,\uparrow,\mathrm{state}}$ and $E_{L,\downarrow,\mathrm{state}}$ of the ZNRs, respectively. For OP-AFM ZGeNR(e) and ZSnNR(f), 
the edge states show anti-crossing of bands with opposite spins and the color of $E_{R,\uparrow,\mathrm{af}}$ changes gradually from red to yellow around the energy gap near $k_y=4\pi/3$, while that of $E_{R,\downarrow,\mathrm{af}}$ changes gradually from yellow to red. For OP-FM ZGeNR(h) and ZSnNR(i), $E_{L,\uparrow,\mathrm{fm}}$ and $E_{L,\downarrow,\mathrm{fm}}$ show anti-crossing of bands with opposite spins near $4\pi/3$. }
\end{figure*}

First, in figures \ref{field}(a)-(c), we show the energy spectra of nonmagnetic 1H/1H ZSiNR, ZGeNR and ZSnNR at $E_z=2E_\mathrm{cr}$. The helical edge states disappear in all these materials. The edge states $E_{R,\uparrow,\mathrm{nm}}$(red) and $E_{R,\downarrow,\mathrm{nm}}$(yellow) [$E_{L,\uparrow,\mathrm{nm}}$(blue) and $E_{L,\downarrow,\mathrm{nm}}$(green)] connect the valence (conduction) band at $k_y=2\pi/3$ and the valence (conduction) band at $k_y=4\pi/3$, and the ZNRs become trivial insulators. However, the trivial edge states of ZSiNR and ZGeNR remain in the bulk energy gap, so they remain metallic states. On the other hand, the trivial edge states of ZSnNR have gap openings different from ZSiNR and ZGeNR, and ZSnNR becomes a trivial insulator. 

Second, in figures \ref{field}(d)-(f), we show the energy spectra of OP-AFM 1H/1H ZSiNR, ZGeNR and ZSnNR. The OP-AFM ZNRs exhibit band splitting due to the broken inversion symmetry. The energy dispersion of ZSiNR shown in \fref{field}(d) hardly changes, as compared to the case without an electric field shown in \fref{nofield}(d). On the other hand, for ZGeNR [\fref{field}(e)] and ZSnNR [\fref{field}(f)], $E_{R,\uparrow,\mathrm{af}}$ (red) and $E_{R,\downarrow,\mathrm{af}}$ (yellow) [$E_{L,\uparrow,\mathrm{af},}$ (blue) and $E_{L,\downarrow,\mathrm{af}}$ (green)] show mixed spin states near $k_y=4\pi/3$. If there are no mixed spin states, $E_{R,\uparrow,\mathrm{af}}$ ($E_{L,\uparrow,\mathrm{af}}$) completely connect the valence (conduction) band at $k_y=2\pi/3$ and the valence (conduction) band at $k_y=4\pi/3$ and $E_{R,\downarrow,\mathrm{af}}$ ($E_{L,\downarrow,\mathrm{af}}$) connect the conduction (valence) band at $k_y=2\pi/3$ and the valence (conduction) band at $k_y=4\pi/3$, as shown in figures \ref{field}(e) and (f).  For the entire  Brillouin zone, OP-AFM ZSiNR is a trivial insulator, while OP-AFM ZGeNR and ZSnNR are metallic. 
 
\begin{figure}[htbp]
\includegraphics[width=8cm]{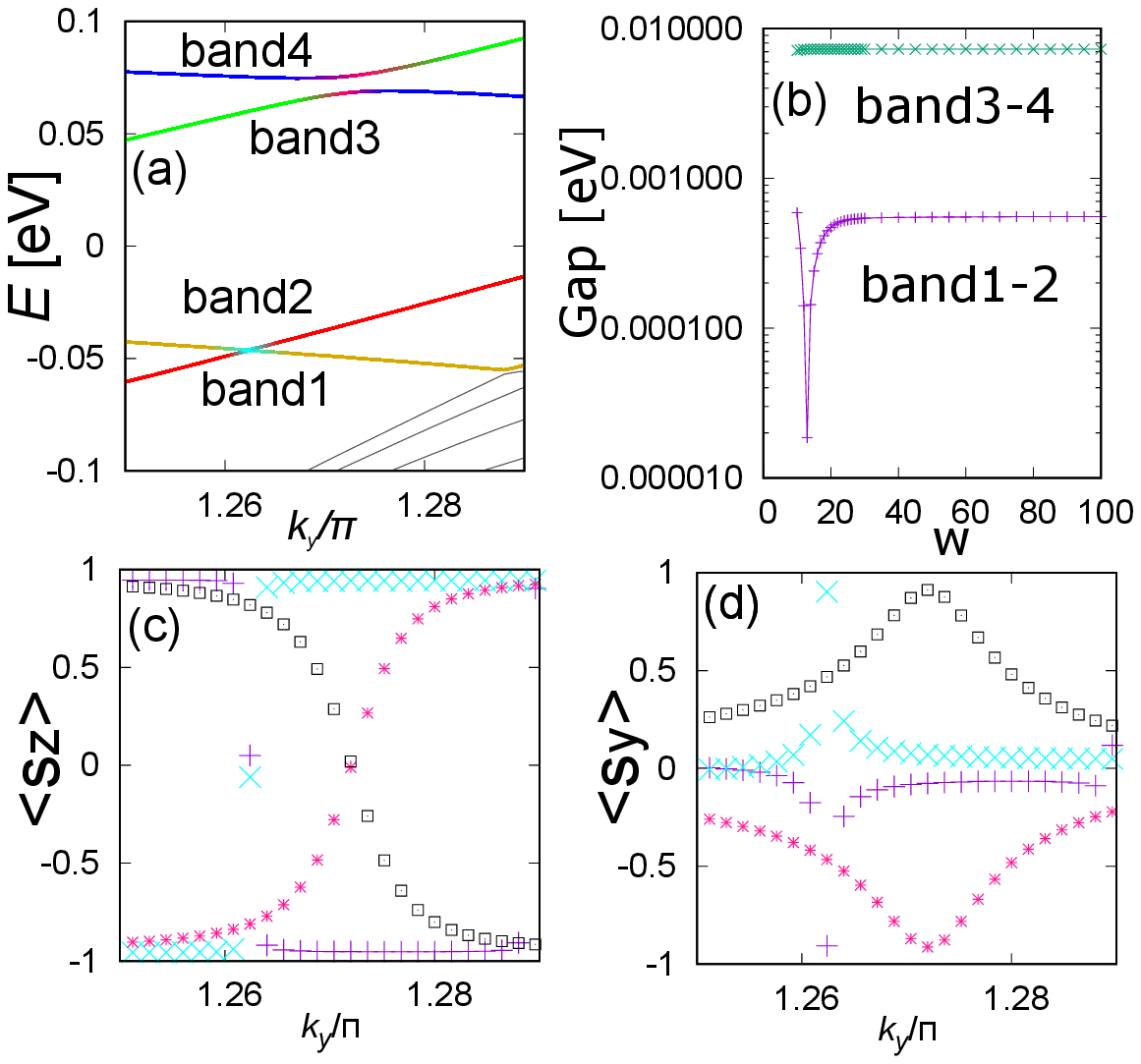}
\caption{\label{invert2} (color online) (a) Energy spectra of OP-AFM ZSnNR at $E_z=2E_\mathrm{cr}^\mathrm{(Sn)}$ [enlarged from \fref{field}(f)]. (b) Energy gap between band1 and band2 and between band3 and band4 as a function of the ribbon width $w$. (c) and (d): The expectation values $\langle s_z\rangle$ and $\langle s_y\rangle$ of the $z$ and $y$ components of the momentum-decomposed spin for band1 (purple), band2 (sky blue), band3 (pink) and band4 (black). }
\end{figure}

Next, we take a closer look at the mixed spin states of OP-AFM 1H/1H ZSnNR at $E_z=2E_\mathrm{cr}^\mathrm{(Sn)}$. \Fref{invert2}(a) shows an  enlargement of the energy spectra of OP-AFM ZSnNR from \fref{field}(f). \Fref{invert2}(b) shows the energy gaps between band1 and band2 and between band3 and band4. 
The energy gap between band1 and band2 becomes smallest at $w=13$ [see \fref{invert2}(b)], and the energy gap increases as the ribbon width increases,  becoming constant ($E_\mathrm{band1-2,gap}\sim 0.000555$ eV) when the ribbon width exceeds 30. 
The energy gap between band3 and band4 becomes constant ($E_\mathrm{band3-4,gap}\sim 0.00728$ eV) for $w\geq 10$. 
Figures \ref{invert2}(c) and (d) show the expectation values $\langle s_z\rangle$ and $\langle s_y\rangle$ of the $z$ and $y$ components of the momentum-decomposed spin of band1 (purple), band2 (sky blue), band3 (pink) and band4 (black). 
The energy gap between band1 and band2 (band3 and band4) inverts near $k_y=1.26\pi$ ($k_y=1.27\pi$) at $w=100$. The difference between the value of $\langle s_y\rangle$ for band1 and that for band2 changes drastically near $k_y=1.26\pi$, while the difference between the value of $\langle s_y\rangle$ for band3 and that for band4 is enhanced around $k_y=1.27\pi$. Band1 and band2 (band3 and band4) show mixed spin states, but they still exhibit anti-crossing of bands with opposite spins even in the presence of an electric field.

Third, we consider the energy spectra of OP-FM 1H/1H ZSiNR, ZGeNR and ZSnNR as shown in figures \ref{field}(g)-(i). 
As compared to OP-AFM ZNRs at $E_z=2E_\mathrm{cr}$ [\ref{field}(d)-(f)], the energy dispersion of the edge states changes, except for the blue line representing $E_{L,\uparrow,\mathrm{fm}}$, as shown in figures \ref{field}(g)-(i). 
By applying an electric field, the crossing points of ZSiNR above (below) the Fermi energy move to the right (left) from $k_y=\pi$ as shown in \fref{field}(g). For OP-FM ZGeNR [\fref{field}(h)] and OP-FM ZSnNR [\fref{field}(i)], the band splitting becomes larger. The edge states $E_{L,\uparrow,\mathrm{fm}}$ (blue) and $E_{L,\downarrow,\mathrm{fm}}$ (green) still show anti-crossing of bands with opposite spins, and $E_{L,\uparrow,\mathrm{fm}}$ and $E_{L,\downarrow,\mathrm{fm}}$ completely connect the conduction band at $k_y=2\pi/3$ and the conduction band at $k_y=4\pi/3$ if there is no anti-crossing of bands, while $E_{R,\uparrow,\mathrm{fm}}$ (red) and $E_{R,\downarrow,\mathrm{fm}}$ (yellow) connect the valence band at $k_y=2\pi/3$ and the valence band at $k_y=4\pi/3$.   
The energy gaps of OP-FM ZSiNR open slightly while the edge states of OP-FM ZGeNR and ZSnNR are metallic. 

The energy spectra of 2H/2H ZNRs are shown in \fref{field_k} (Appendix). The 2H/2H ZNRs show similar behavior to the 1H/1H ZNRs. 


Finally, in order to understand the electric field effect within the effective low energy model, we add $\mathcal{H}_\mathrm{v} = E_z\sigma_0\tau_z$ to the low-energy effective model [equation \eref{eff}]. $\mathcal{H}_\mathrm{v}$ describes the staggered potential that breaks the inversion symmetry and drives spin splitting of the edge states. Thus, we can qualitatively explain the energy dispersion of the 1H/1H ZNRs in the presence of a vertical electric field. 

\begin{figure*}[htbp]
\includegraphics[width=16cm]{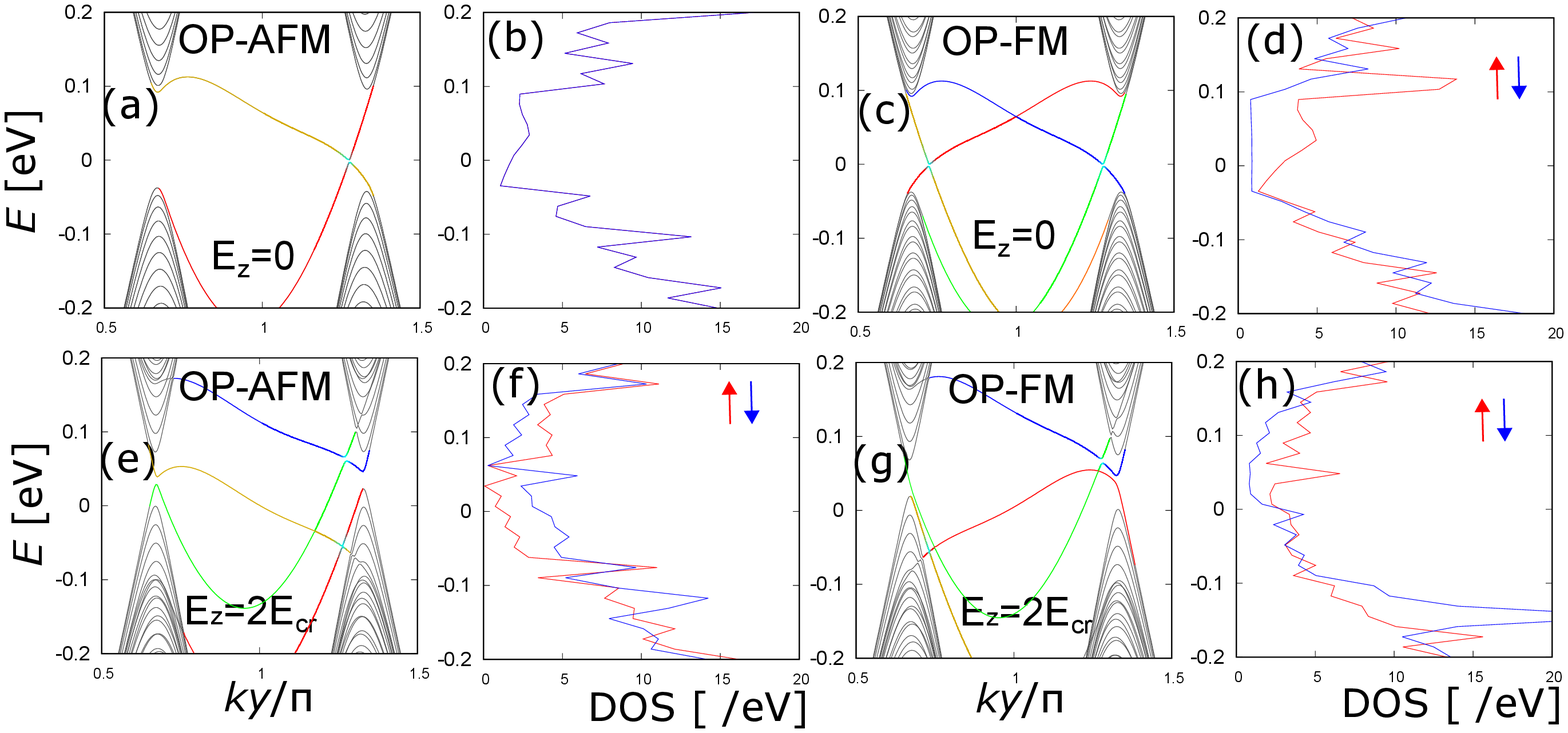}
\caption{\label{afmfmdos} (color online) (a) and (e): The energy spectra of OP-AFM 1H/1H ZSnNR with $E_z=0$ and $2E^\mathrm{(Sn)}_\mathrm{cr}$. (b) and (f): The corresponding spin-resolved densities of states. (c) and (g): The energy spectra of OP-FM 1H/1H ZSnNR with $E_z=0$ and  $2E^\mathrm{(Sn)}_\mathrm{cr}$. (d) and (h): The corresponding spin-resolved densities of states. }
\end{figure*}

It has been proposed that ZNRs can become topological quantum field effect transistor (TQFET) controlling the helical edge state by applying the external electric field \cite{Ezawa_externalfield}. Indeed, it is indicated in Ref. \cite{Ezawa_externalfield} that ZNRs become trivial insulators at $E_z=2E_\mathrm{cr}$. However, it is not obvious how to apply this result to actual materials, because the energy dispersion of the actual edge states of ZNRs shows complicated behavior and a metal-insulator transition of the edge states cannot occur easily. Our results indicate instead that it will be difficult to realize TQFET using ZSiNR, ZGeNR or ZSnNR,  
although they can be used as spin-polarizers. In figures \ref{afmfmdos}(b), (f), (d) and (h), we show the spin-resolved densities of states for OP-AFM 1H/1H ZSnNR with $E_z=0$ and $2E^\mathrm{(Sn)}_\mathrm{cr}$ and for OP-FM 1H/1H ZSnNR with $E_z=0$ and $2E^\mathrm{(Sn)}_\mathrm{cr}$. For OP-AFM ZSnNR without an electric field, spin-polarization does not occur, but a large spin-polarization is switched on by an electric field. We also find that in OP-FM ZSnNR the qualitative behavior of the spin-polarization is changed by the electric field. These features can be applied for a spin-polarizer. 

\section{Discussion and Summary}\label{sec4} 
We have studied the energy spectra of the spin-resolved edge states of ZSiNR, ZGeNR, and ZSnNR with and without a vertical electric field based on a multi-orbital tight-binding model with nonmagnetic, OP-AFM, and OP-FM at the edges. We have found an effective low-energy model for non-linear dispersion of the edge states of the ZNRs. The effective low-energy model can explain the edge states qualitatively. We find that the edge states of ZNRs with large spin-orbit coupling show anti-crossing of bands with opposite spins, even if only out-of-plane edge magnetization exists. They also show a variety of edge states; however, the realization of a field effect transistor--as proposed on the basis of a single-orbital tight-binding model--is not easy due to the non-linear energy dispersion of the edge states. 

On the other hand, our study of the spin-resolved local densities of states of OP-AFM and OP-FM ZSnNRs has shown that the spin-polarization at the edge states can be controlled by an applied electric field. This property can serve as a guide for the fabrication of spin-polarizers based on ZNRs. 

Finally, we have compared our results with those of the effective single-orbital model described in Ref. \cite{systematic_stanene}. By adding new terms, Ref. \cite{systematic_stanene} shows that the single-orbital model can reproduce the non-linear dispersion of edge states. However, the model cannot explain anti-crossing of bands with opposite spins for edge states with out-of-plane edge magnetization. 
We also find that--in the multi-orbital model--the value of $\langle s_y\rangle$ for the edge states becomes non-zero in the bulk energy gap even if we consider the magnetization along the z-direction. The on-site spin-orbit coupling $\vec{L}\cdot\vec{s}$ produces such magnetic anisotropy. Therefore, the multi-orbital tight-binding model is indispensable for understanding the edge states of ZSiNR, ZGeNR, and ZSnNR with edge magnetization.

\ack
This work was supported by a Grant-in Aid for Scientific Research on Innovative Areas `Topological Material Science' (Grant No. JP15H05855,JP15H05853), a Grant-in-Aid for Challenging Exploratory Research (Grant No. JP15K13498,JP18K19020), the Core Research for Evolutional Science and Technology  (CREST) of the Japan Science and Technology Corporation (JST) (Grant No. JPMJCR14F1). 

\section*{Appendix}
\subsection*{Energy spectra of 2H/2H ZNRs}
\begin{figure*}[htbp]
\includegraphics[width=17cm]{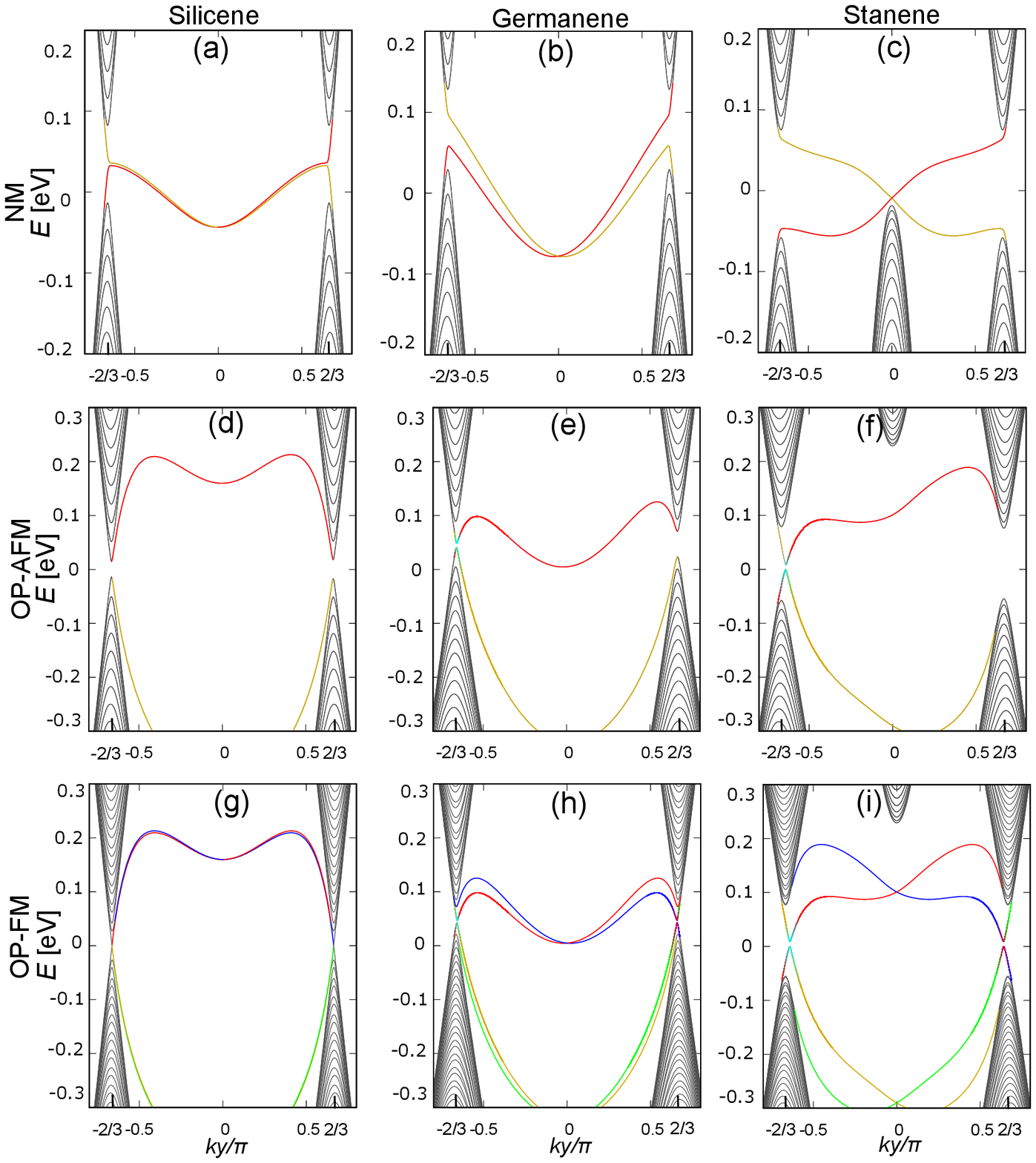}
\caption{\label{nofield_k} (color online) Energy spectra of nonmagnetic (a)-(c), OP-AFM (d)-(f) and OP-FM (g)-(i) 2H/2H ZSiNRs, ZGeNRs, and ZSnNRs  with $w=100$. The value $E=0$ represents the Fermi energy. Red, yellow, blue and green denote the edge states $E_{R,\uparrow,\mathrm{state}}$, $E_{R,\downarrow,\mathrm{state}}$, $E_{L,\uparrow,\mathrm{state}}$ and $E_{L,\downarrow,\mathrm{state}}$ of the ZNRs, respectively. For OP-AFM ZGeNR (e) and ZSnNR (f), 
the edge states show anti-crossing of bands with opposite spins, and the color of $E_{R,\uparrow,\mathrm{af}}$ gradually changes from red to yellow around the energy gap near $k_y=-2\pi/3$, while that of $E_{R,\downarrow,\mathrm{af}}$ gradually changes from yellow to red. Also, for OP-FM ZGeNR (h) and ZSnNR (i), $E_{R,\uparrow,\mathrm{fm}}$ and $E_{R,\downarrow,\mathrm{fm}}$ ($E_{L,\uparrow,\mathrm{fm}}$ and $E_{L,\downarrow,\mathrm{fm}}$) show anti-crossing of bands with opposite spins near $k_y=-2\pi/3$ ($2\pi/3$).}
\end{figure*}

\begin{figure*}[htbp]
\includegraphics[width=17cm]{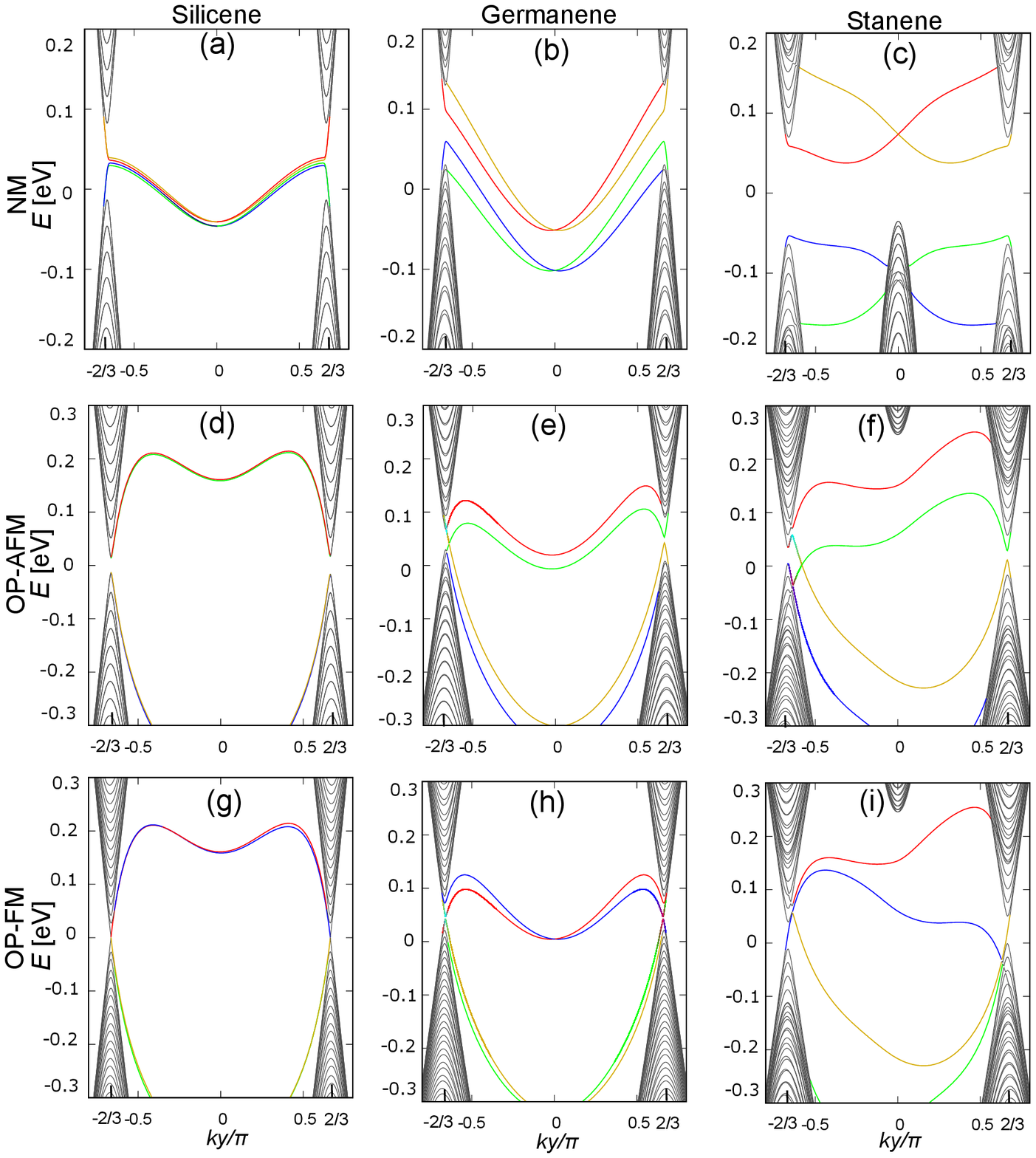}
\caption{\label{field_k} (color online) Energy spectra of nonmagnetic (a)-(c), OP-AFM (d)-(f) and OP-FM (g)-(i) 2H/2H ZSiNRs, ZGeNRs and ZSnNRs in the presence of a vertical electric field $E_z=2E_\mathrm{cr}$  for $w=100$. The value $E=0$ represents the Fermi energy. Red, yellow, blue and green denote $E_{R,\uparrow,\mathrm{state}}$, $E_{R,\downarrow,\mathrm{state}}$, $E_{L,\uparrow,\mathrm{state}}$ and $E_{L,\downarrow,\mathrm{state}}$ of ZNRs, respectively. }
\end{figure*}

Here, we discuss the energy spectra of nonmagnetic, OP-AFM and OP-FM 2H/2H ZNRs without and with a vertical electric field, as shown in figures \ref{nofield_k} and \ref{field_k}. In all states of 2H/2H ZNRs, the edge states appear in the range $-2\pi/3 \leq k_y \leq 2\pi/3$. 
Here, we identify the energy dispersion of the edge states with up spins at a right edge (red line), those with down spins at a right edge (yellow line), those with up spins at a left edge (blue line) and those with down spins at a left edge (green line) as the $E_{R,\uparrow,\mathrm{state}}$, $E_{R,\downarrow,\mathrm{state}}$, $E_{L,\uparrow,\mathrm{state}}$, and $E_{L,\downarrow,\mathrm{state}}$, respectively. 
Similar to the 1H/1H ZNRs, the subscript `state' denotes either the nonmagnetic case (nm), the OP-AFM case (af) or the OP-FM case (fm).

The energy spectra of nonmagnetic ZSiNR, ZGeNR and ZSnNR are plotted in figures \ref{nofield_k}(a)-(c), respectively. They each show helical edge states. The edge states $E_\mathrm{R,\uparrow,nm}$ (red) connects the valence band at $k_y=-2\pi/3$ and the conduction band at $k_y=2\pi/3$;  $E_\mathrm{R,\downarrow,nm}$ (yellow) connects the conduction band at $k_y=-2\pi/3$ and the valence band at $k_y=2\pi/3$; $E_\mathrm{R,\uparrow,nm}$ and $E_\mathrm{R,\downarrow,nm}$ cross at $k_y=0$. $E_\mathrm{R,\uparrow,nm}$ and $E_\mathrm{L,\downarrow,nm}$ (green) [$E_\mathrm{R,\downarrow,nm}$ and $E_\mathrm{L,\uparrow,nm}$ (blue)] are degenerate. Both ZSiNR and ZGeNR show downward-convex dispersion, while ZSnNR shows linear dispersion near $k_y=0$. 

Figures \ref{nofield_k}(d)-(f) show the energy spectra of OP-AFM 2H/2H ZSiNR, ZGeNR and ZSnNR. 
Although time-reversal and inversion symmetry are broken by the magnetization, the combined symmetry remains.
Therefore, $E_{R,\uparrow,\mathrm{af}}$ (red) and $E_{L,\downarrow,\mathrm{af}}$ (green) [$E_{R,\downarrow,\mathrm{af}}$ (yellow) and $E_{L,\uparrow,\mathrm{af}}$ (blue)] are doubly degenerate.  
\Fref{nofield_k}(d) shows that the edge states of OP-AFM ZSiNR have energy gaps due to the edge magnetization and 
that the energy gap at $k_y=-2\pi/3$ is similar in size to that at $k_y=2\pi/3$. 
On the other hand, for OP-AFM 2H/2H ZGeNR [\fref{nofield_k}(e)] and ZSnNR [\fref{nofield_k}(f)], the energy dispersion has an energy gap at $k_y=2\pi/3$ and almost crosses at $k_y=-2\pi/3$. The edge states $E_\mathrm{R,\uparrow,af}$ and $E_\mathrm{R,\downarrow,af}$ show mixed spin states, as is also the case for 1H/1H OP-AFM ZGeNR and ZSnNR [figures \ref{nofield}(e) and (f)], and the color of $E_{R,\uparrow,\mathrm{af}}$ gradually changes from red to yellow around the energy gap near $k_y=-2\pi/3$ as it goes to from left to right along $k_y$, while that of $E_{R,\downarrow,\mathrm{af}}$ changes from yellow to red around the energy gap near $k_y=-2\pi/3$. 

Figures \ref{nofield_k}(g)-(i) show the energy spectra of OP-FM 2H/2H ZSiNR, ZGeNR and ZSnNR. 
Although time-reversal symmetry is broken by the magnetization, inversion symmetry remains. 
The OP-FM ZNRs thus show symmetric dispersion with respect to $k_y=0$. 
As compared to OP-AFM ZNRs, only the direction of the magnetization around the left side edge is reversed (see figures \ref{sil_mag}, \ref{ger_mag}, \ref{sta_mag}). As a result, $E_{R,\uparrow,\mathrm{fm}}(k_y)$ (red) and $E_{R,\downarrow,\mathrm{fm}}(k_y)$ (yellow) in figures \ref{nofield_k}(g)-(i) coincide with $E_{R,\uparrow,\mathrm{af}}(k_y)$ (red) and $E_{R,\downarrow,\mathrm{af}}(k_y)$ (yellow) in figures \ref{nofield_k}(d)-(f).  
On the other hand, $E_{L,\uparrow,\mathrm{fm}}(k_y)$ (blue) and $E_{L,\downarrow,\mathrm{fm}}(k_y)$ (green) in figures \ref{nofield}(g)-(i) coincide with $E_{R,\uparrow,\mathrm{af}}(-k_y)$ (red) and $E_{R,\downarrow,\mathrm{af}}(-k_y)$ (yellow). 
The energy gaps of OP-FM ZSiNR are almost zero. 
On the other hand, for OP-FM ZGeNR [\fref{nofield_k}(h)] and ZSnNR [\fref{nofield_k}(i)], $E_\mathrm{R,\uparrow,fm}$ and $E_\mathrm{R,\downarrow,fm}$ show mixed spin states near $k_y=-2\pi/3$ and the spins of $E_\mathrm{L,\uparrow,fm}$ and $E_\mathrm{L,\downarrow,fm}$ also flip near $k_y=2\pi/3$.

Next, in \fref{field_k} we show the energy spectra of nonmagnetic, OP-AFM and OP-FM ZNRs with $E_z=2E_\mathrm{cr}$.  
Figures \ref{field_k}(a)-(c) show the energy spectra of nonmagnetic ZSiNR, ZGeNR and ZSnNR. The helical edge states disappear in the bulk energy gap. The edge states $E_\mathrm{R,\uparrow,nm}$ (red) and $E_\mathrm{R,\downarrow,nm}$ (yellow) connect the conduction band at $k_y=-2\pi/3$ and the conduction band at $k_y=2\pi/3$, while $E_\mathrm{L,\uparrow,nm}$ (blue) and $E_\mathrm{L,\downarrow,nm}$ (green) connect the valence band at $k_y=-2\pi/3$ and the valence band at $k_y=2\pi/3$. These ZNRs thus become trivial insulators. However, the trivial edge states of ZSiNR and ZGeNR remain in the bulk energy gap, so they remain metallic states. On the other hand, the trivial edge states of ZSnNR have a gap opening, and ZSnNR becomes a trivial insulator. 

Figures \ref{field_k}(d)-(f) show the energy spectra of OP-AFM ZSiNR, ZGeNR, and, ZSnNR with $E_z=2E_\mathrm{cr}$. For OP-AFM ZSiNR [\fref{field_k}(d)],  
the energy dispersion hardly changes as compared to that without an electric field, as shown in \fref{nofield_k}(d). 
Thus, OP-AFM ZSiNR with $E_z=2E^\mathrm{(Si)}_\mathrm{cr}$ remains in an insulating state. On the other hand, for OP-AFM ZGeNR [\fref{field_k}(e)]  and ZSnNR [\fref{field_k}(f)], $E_\mathrm{R,\uparrow,af}$ (red) [$E_\mathrm{L,\uparrow,af}$ (blue)] connect the conduction (valence) band at $k_y=-2\pi/3$ and the conduction (valence) band at $k_y=2\pi/3$, while $E_\mathrm{R,\downarrow,af}$ (yellow) [$E_\mathrm{L,\downarrow,af}$ (green)]  connect the conduction (valence) band at $k_y=-2\pi/3$ and the valence (conduction) band at $k_y=2\pi/3$. For the entire Brillouin zone, OP-AFM ZSiNR with $E_z=2E_\mathrm{cr}$ is a trivial insulator, while the edge states of ZGeNR and ZSnNR with $E_z=2E_\mathrm{cr}$ are metallic. 

Figures \ref{field_k}(g)-(i) show the energy spectra of OP-FM ZSiNR, ZGeNR and ZSnNR. The energy spectra of OP-FM ZSiNR are insensitive to an electric field as shown in \fref{field_k}(g). For OP-FM ZGeNR [\fref{field_k}(h)] and ZSnNR [\fref{field_k}(i)], $E_\mathrm{R,\uparrow,fm}$ (red) and $E_\mathrm{R,\downarrow,fm}$ (yellow) connect the conduction band at $k_y=-2\pi/3$ and the conduction band at $k_y=2\pi/3$, while $E_\mathrm{L,\uparrow,fm}$ (blue) and $E_\mathrm{L,\downarrow,fm}$ (green) connect the valence band at $k_y=-2\pi/3$ and the valence band at $k_y=2\pi/3$. For the entire Brillouin zone, OP-FM ZNRs with $E_z=2E_\mathrm{cr}$ are metallic. 


\subsection*{The expectation value of the $z$ component of spin}
\begin{figure}[htbp]
\includegraphics[width=7cm]{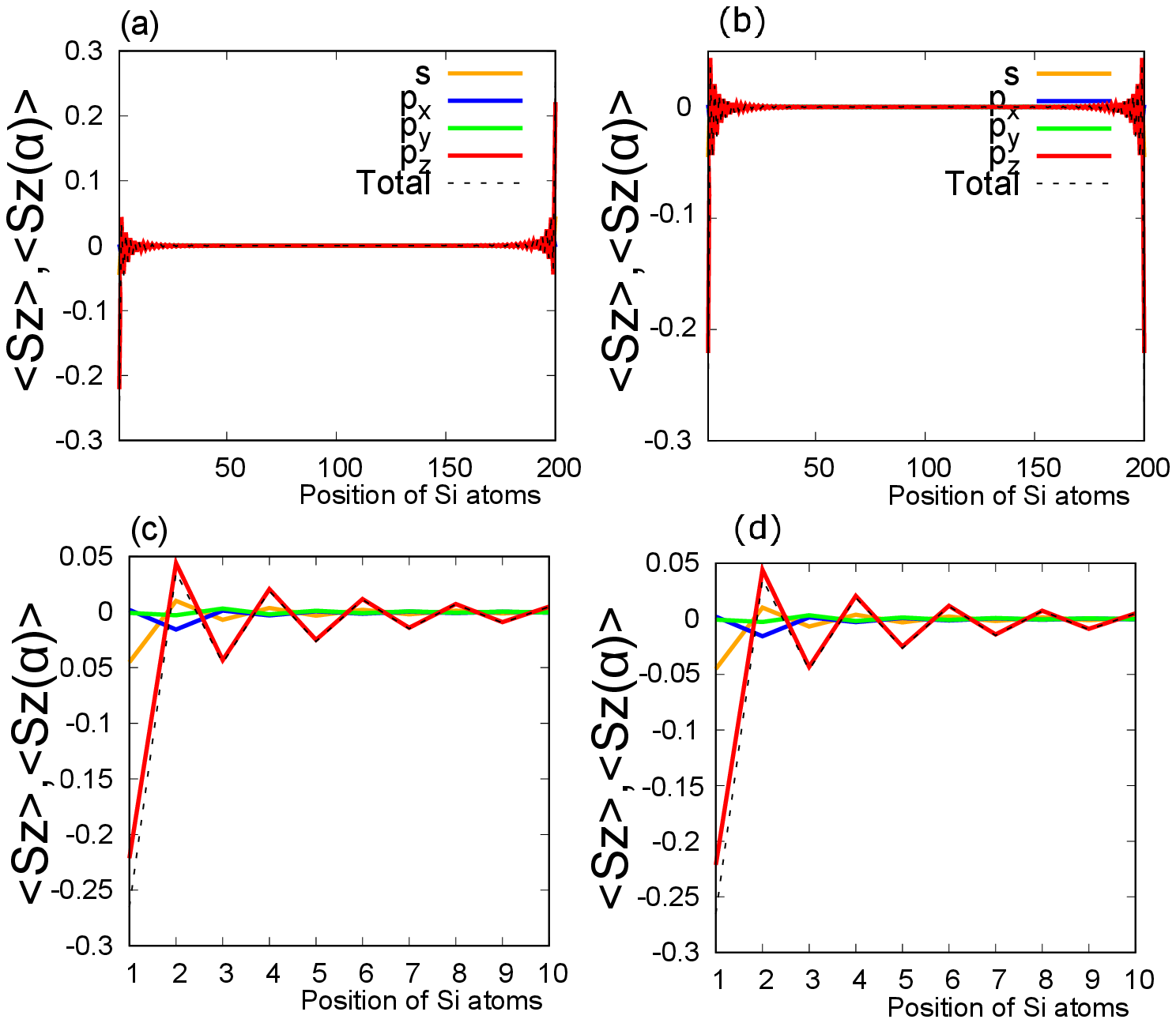}
\caption{\label{sil_mag} (color online) $\langle S_z\rangle$ and $\langle S_z(\alpha)\rangle$ for OP-AFM [(a) and (c)] and OP-FM [(b) and (d)] 1H/1H ZSiNRs. }
\end{figure}

\begin{figure}[htbp]
\includegraphics[width=7cm]{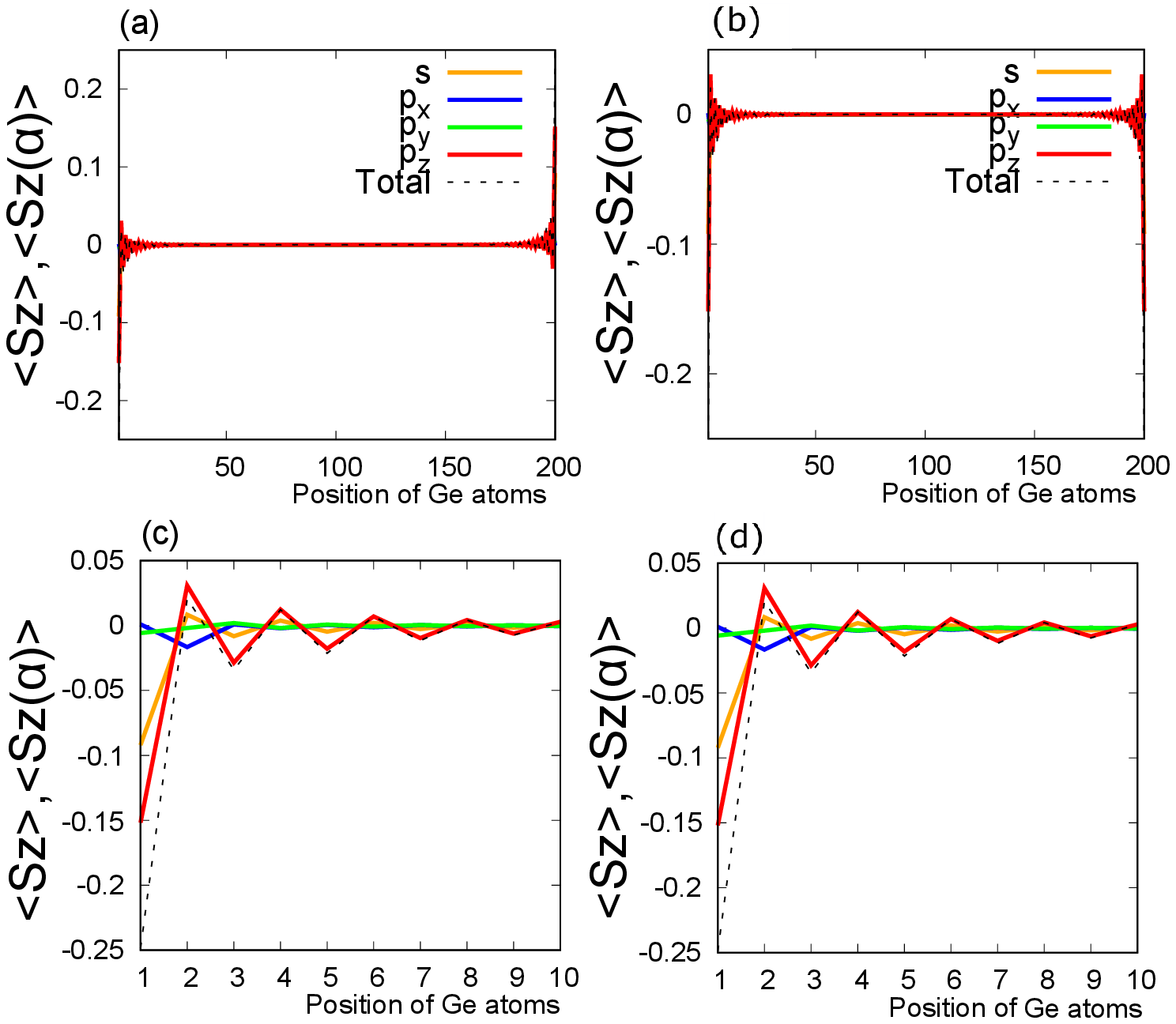}
\caption{\label{ger_mag} (color online) $\langle S_z\rangle$ and $\langle S_z(\alpha)\rangle$ for OP-AFM [(a) and (c)] and OP-FM [(b) and (d)] 1H/1H ZGeNRs. }
\end{figure}

\begin{figure}[htbp]
\includegraphics[width=7cm]{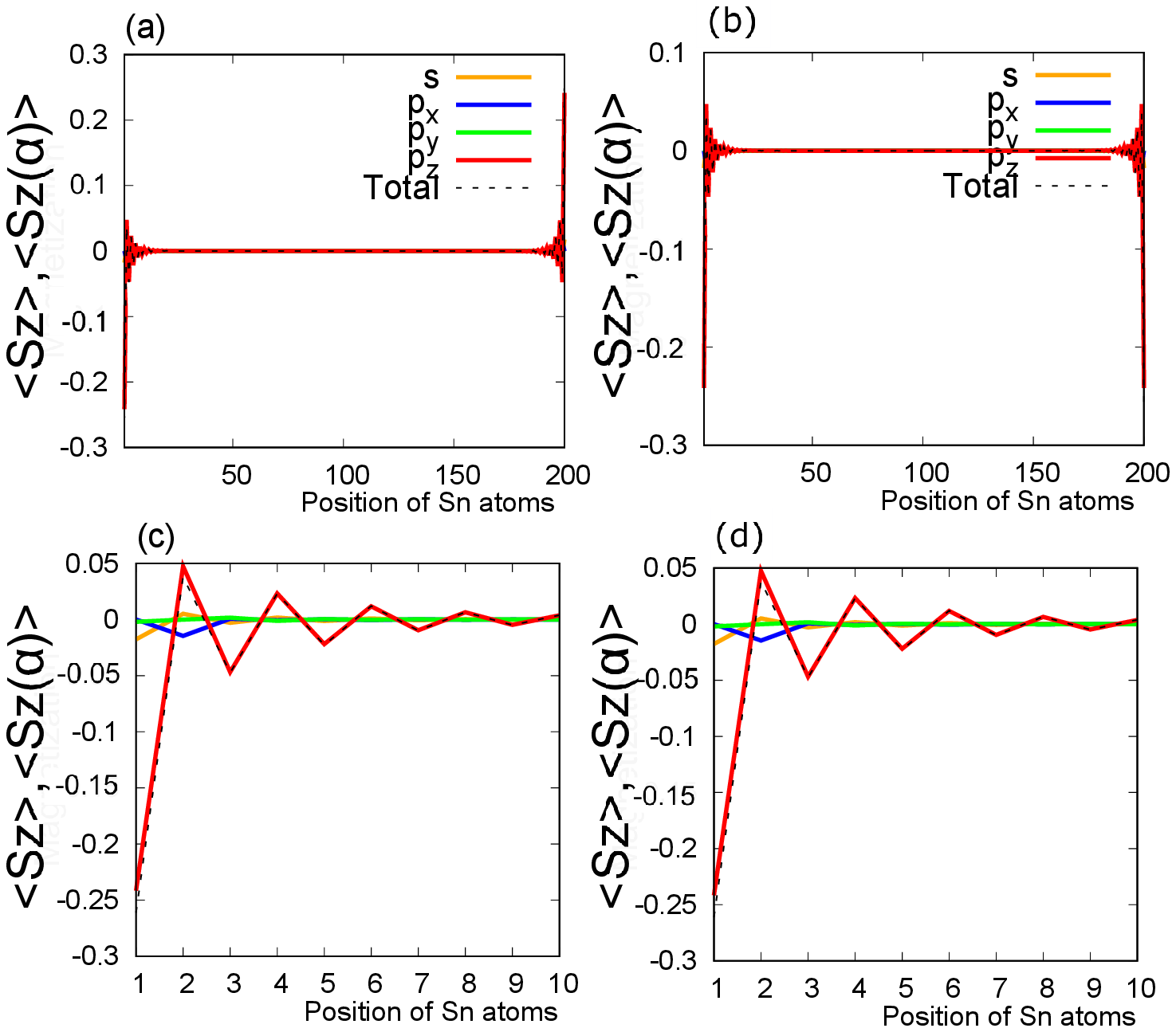}
\caption{\label{sta_mag} (color online) $\langle S_z\rangle$ and $\langle S_z(\alpha)\rangle$ for OP-AFM [(a) and (c)] and OP-FM [(b) and (d)] 1H/1H ZSnNRs. }
\end{figure}

Here, we discuss the expectation value $\langle S_z\rangle$ of the $z$ component of the spin for 1H/1H OP-AFM and OP-FM ZNRs without an electric field as a function of the position of the silicon, germanium or tin atoms. The OP-AFM ZNRs have ferromagnetic order at the edge sites, and the spin directions at the left and right edges are opposite to each other. The absolute value of $\langle S_z\rangle$ decreases exponentially from the edge sites into the bulk, and the spin directions adjacent to each other are opposites. The magnitude of $\langle S_z\rangle$ of 1H/1H ZNRs is largest at most edge sites, while that of 2H/2H ZNRs is largest at the position adjacent to most edge sites. We next consider the orbital decompositions $\langle S_z(\alpha)\rangle$, where $\alpha$ represents $s$, $p_x$, $p_y$ or $p_z$ orbitals. Then, 
\begin{eqnarray}
\langle S_z\rangle = \sum_\alpha \langle S_z(\alpha)\rangle
\end{eqnarray}
The absolute value of $\langle S_z\rangle$ for both OP-AFM and OP-FM 1H/1H ZSiNRs in \fref{sil_mag} is 0.265. The component $\langle S_z(p_z)\rangle$ is most dominant and $\langle S_z(s)\rangle$ is next-most dominant. Further, $\langle S_z(p_x)\rangle$ is largest at positions adjacent to  most edge sites, while $\langle S_z(p_y)\rangle$ is almost zero. 
The value of $\langle S_z\rangle$ for both OP-AFM and OP-FM 1H/1H ZGeNRs in \fref{ger_mag} is 0.248. Again, the component $\langle S_z(p_z)\rangle$ is most dominant, and the magnitude of $\langle S_z(s)\rangle$ is two-thirds of $\langle S_z(p_z)\rangle$. The component $\langle S_z(p_x)\rangle$ is the largest adjacent to most edge sites, while $\langle S_z(p_y)\rangle$ is almost zero. 
The absolute value of $\langle S_z\rangle$ for both OP-AFM and OP-FM 1H/1H ZSnNRs in \fref{sta_mag} is 0.261. The spatial dependences of $\langle S_z\rangle$ and $\langle S_z(\alpha)\rangle$ in ZSnNRs are similar to those of ZSiNRs. 

\begin{figure}[htbp]
\includegraphics[width=7cm]{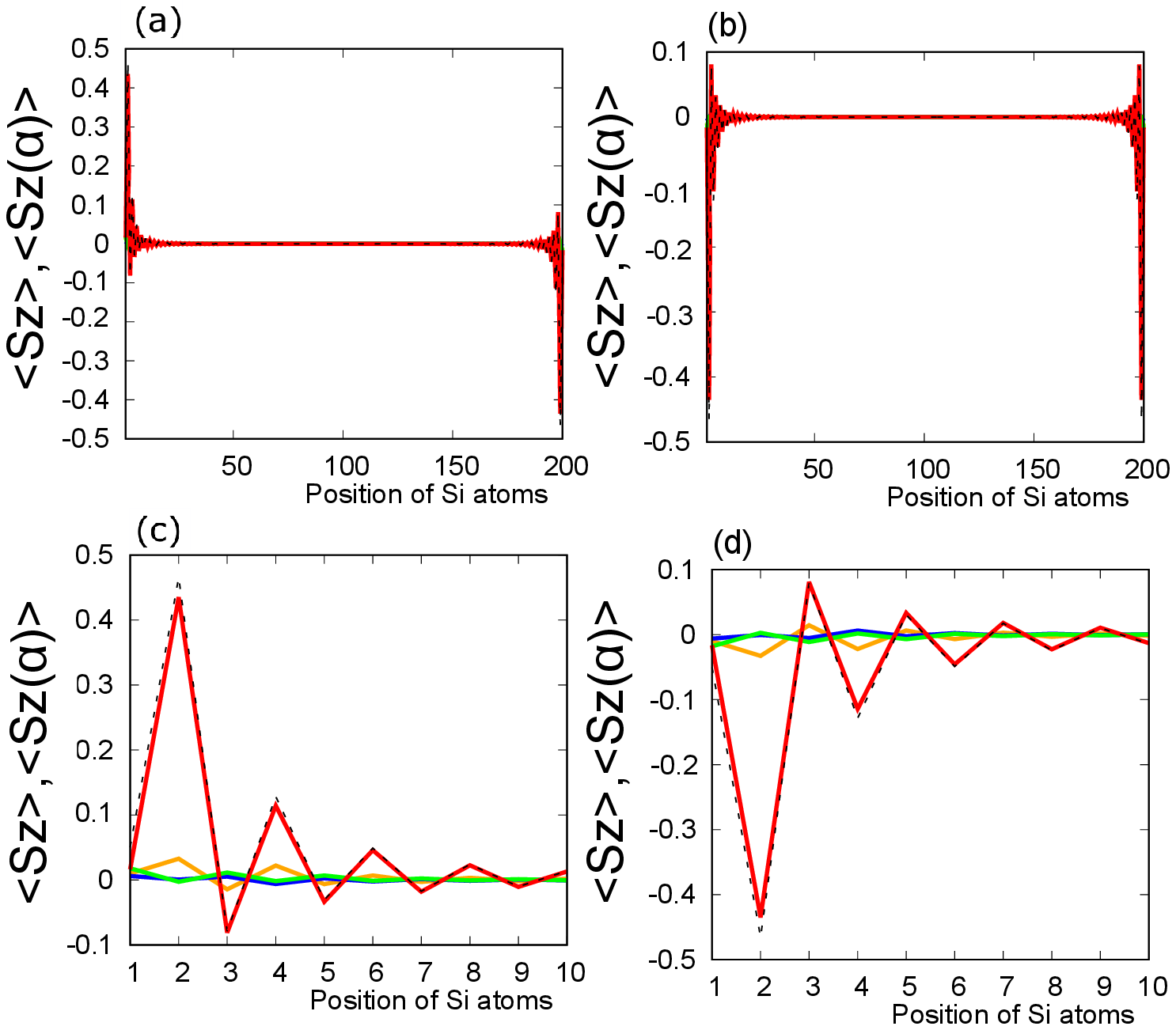}
\caption{\label{sil_mag_k} (color online) $\langle S_z\rangle$ and $\langle S_z(\alpha)\rangle$ of OP-AFM (a)(c) and OP-FM (b)(d)2H/2H ZSiNRs. }
\end{figure}

\begin{figure}[htbp]
\includegraphics[width=7cm]{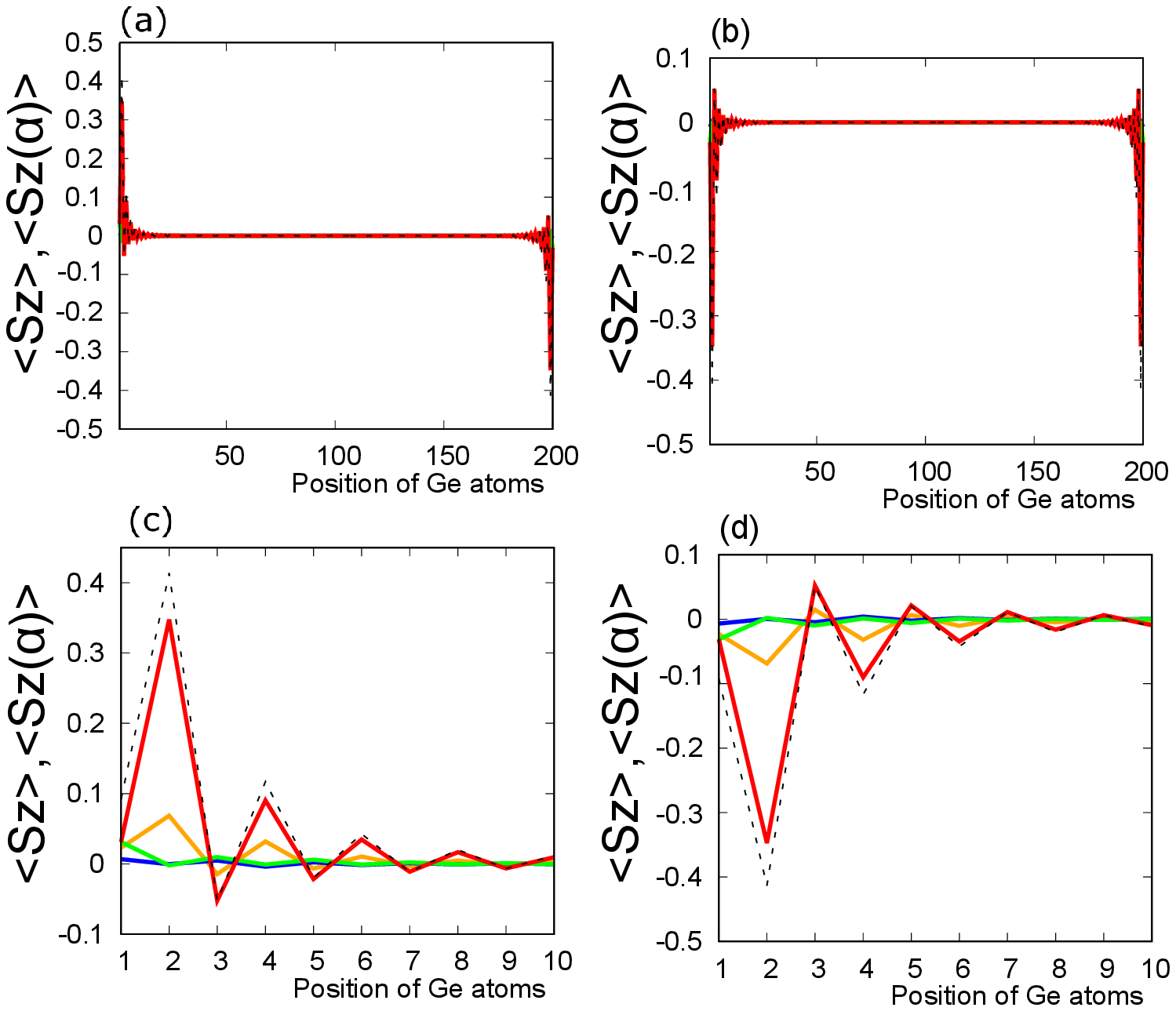}
\caption{\label{ger_mag_k} (color online) $\langle S_z\rangle$ and $\langle S_z(\alpha)\rangle$ of OP-AFM (a)(c) and OP-FM (b)(d)2H/2H ZGeNRs. }
\end{figure}

\begin{figure}[htbp]
\includegraphics[width=7cm]{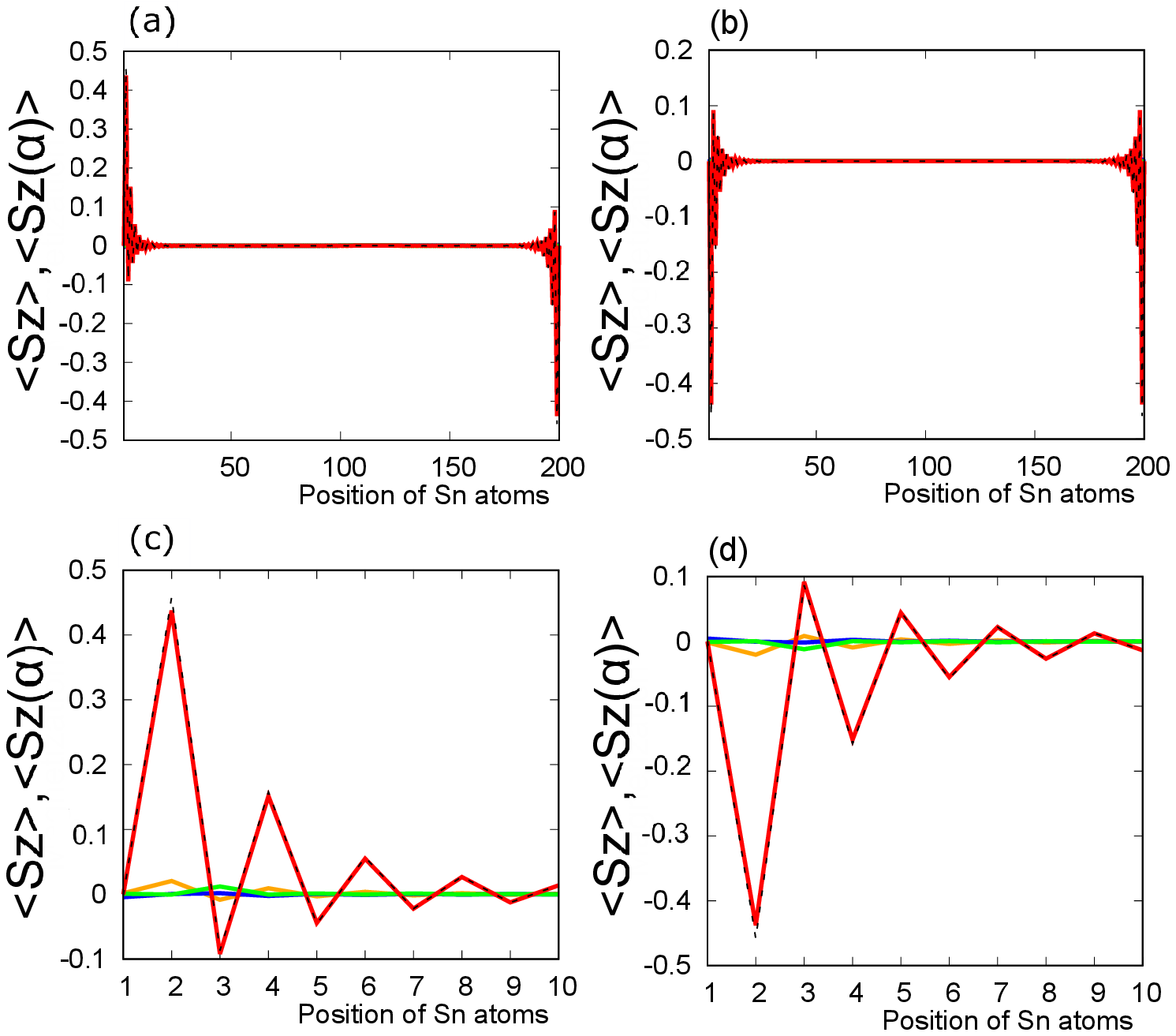}
\caption{\label{sta_mag_k} (color online) $\langle S_z\rangle$ and $\langle S_z(\alpha)\rangle$ of OP-AFM (a)(c) and OP-FM (b)(d)2H/2H ZSnNRs. }
\end{figure}

Next, we consider $\langle S_z\rangle$ for 2H/2H OP-AFM and OP-FM ZNRs without an electric field (see figures \fref{sil_mag_k}, \ref{ger_mag_k} and \ref{sta_mag_k}). The OP-AFM has ferromagnetic order at the edge sites, and the spin directions at the left and right edges are opposites. The absolute value of $\langle S_z\rangle$ decreases exponentially from the edge sites into the bulk, and adjacent spin directions are opposites. The magnitude of $\langle S_z\rangle$ for 2H/2H ZNRs is the largest adjacent to most edge sites. 
The absolute value of $\langle S_z\rangle$ for both OP-AFM and OP-FM 2H/2H ZSiNRs in \fref{sil_mag_k} is 0.466. The component $\langle S_z(p_z)\rangle$ is most dominant, and $\langle S_z(s)\rangle$ is the next-most dominant. The magnitude of $\langle S_z(p_y)\rangle$ is small and decays with oscillations, while $\langle S_z(p_x)\rangle$ is almost zero. 
The absolute value of $\langle S_z\rangle$ for both OP-AFM and OP-FM 2H/2H ZGeNRs in \fref{ger_mag_k} is 0.414. The component $\langle S_z(p_z)\rangle$ is most dominant and $\langle S_z(s)\rangle$ is the next-most dominant. The component $\langle S_z(p_y)\rangle$ is small and decays with oscillations, while $\langle S_z(p_x)\rangle$ is almost zero.
The absolute value of $\langle S_z\rangle$ for both OP-AFM and OP-FM 2H/2H ZSnNRs in \fref{sta_mag_k} is 0.458. The spatial dependences of $\langle S_z\rangle$ and $\langle S_z(\alpha)\rangle$ for ZSnNRs are similar to those for ZSiNRs. 

\section*{Reference}
\providecommand{\newblock}{}


\end{document}